\documentclass[english,aps,reprint,showpacs,prb,amsmath,amssymb,floatfix,superscriptaddress,eqsecnum]{revtex4-1}
\usepackage[T1]{fontenc}
\usepackage[latin9]{inputenc}
\usepackage{babel}
\usepackage{textcomp}
\usepackage{amsmath}
\usepackage{amssymb}
\usepackage{graphicx}
\usepackage{wasysym}
\usepackage[unicode=true,
 bookmarks=true,bookmarksnumbered=false,bookmarksopen=false,
 breaklinks=false,pdfborder={0 0 1},backref=false,colorlinks=false]
 {hyperref}
\hypersetup{
 pdfauthor={Simon Streib}}

\makeatletter
\bibpunct{[}{]}{,}{n}{}{}
\usepackage{placeins}

\makeatother

\begin{document}

\title{Magnon-phonon interactions in magnetic insulators}

\author{Simon Streib}

\affiliation{Kavli Institute of NanoScience, Delft University of Technology, Lorentzweg
1, 2628 CJ Delft, The Netherlands}

\author{Nicolas Vidal-Silva}

\affiliation{Departamento de F\'{\i}sica, Universidad de Santiago de Chile, Avda. Ecuador
3493, Santiago, Chile}

\affiliation{Center for the Development of Nanoscience and Nanotechnology (CEDENNA),
917-0124 Santiago, Chile}

\affiliation{Departamento de Física, Facultad de Ciencias Físicas y Matemáticas,
Universidad de Chile, Casilla 487-3, Santiago, Chile}

\author{Ka Shen}

\affiliation{Department of Physics, Beijing Normal University, Beijing 100875,
China}

\author{Gerrit E. W. Bauer}

\affiliation{Kavli Institute of NanoScience, Delft University of Technology, Lorentzweg
1, 2628 CJ Delft, The Netherlands}

\affiliation{Department of Physics, Beijing Normal University, Beijing 100875,
China}

\affiliation{Institute for Materials Research \& WPI-AIMR \& CSRN, Tohoku University,
Sendai 980-8577, Japan}

\date{June 4, 2019}
\begin{abstract}
We address the theory of magnon-phonon interactions and compute the
corresponding quasi-particle and transport lifetimes in magnetic insulators
with focus on yttrium iron garnet at intermediate temperatures from
anisotropy- and exchange-mediated magnon-phonon interactions, the
latter being derived from the volume dependence of the Curie temperature.
We find in general weak effects of phonon scattering on magnon transport
and the Gilbert damping of the macrospin Kittel mode. The magnon transport
lifetime differs from the quasi-particle lifetime at shorter wavelengths.
\end{abstract}
\maketitle

\section{Introduction\label{sec:Introduction}}

Magnons are the elementary excitations of magnetic order, i.e. the
quanta of spin waves. They are bosonic and carry spin angular momentum.
Of particular interest are the magnon transport properties in yttrium
iron garnet (YIG) due to its very low damping ($\alpha<10^{-4}$),
which makes it one of the best materials to study spin-wave or spin
caloritronic phenomena \cite{Serga2010,Kajiwara2010,Cornelissen2016,Bauer2012,Li2016,Cornelissen2015}.
For instance, the spin Seebeck effect (SSE) in YIG has been intensely
studied in the past decade \cite{Uchida2010,Uchida2010b,Kikkawa2013,Siegel2014,Jin2015,Kehlberger2015,Iguchi2017}.
Here, a temperature gradient in the magnetic insulator injects a spin
current into attached Pt contacts that is converted into a transverse
voltage by the inverse spin Hall effect. Most theories explain the
effect by thermally induced magnons and their transport to and through
the interface to Pt \cite{Uchida2010,Xiao2010,Adachi2011,Jaworski2011,Adachi2013,Schreier2013,Rezende2014}.
However, phonons also play an important role in the SSE through their
interactions with magnons \cite{Adachi2010,Uchida2012b,Kikkawa2016}.

Magnetoelastic effects in magnetic insulators were addressed first
by Abrahams and Kittel \cite{Abrahams1952,Kittel1953,Kittel1958},
and by Kaganov and Tsukernik \cite{Kaganov1959}. In the long-wavelength
regime, the strain-induced magnetic anisotropy is the most important
contribution to the magnetoelastic energy, whereas for shorter wavelengths,
the contribution from the strain-dependence of the exchange interaction
becomes significant \cite{Gurevich1996,Rueckriegel2014,Maehrlein2017}.
Rückriegel \emph{et al. }\cite{Rueckriegel2014} computed very small
magnon decay rates in thin YIG films due to magnon-phonon interactions
with quasi-particle lifetimes $\tau_{qp}\apprge480\;\mathrm{ns,}$even
at room temperature. However, these authors do not consider the exchange
interaction and the difference between quasi-particle and transport
lifetimes.

Recently, it has been suggested that magnon spin transport in YIG
at room temperature is driven by the magnon chemical potential \cite{Cornelissen2016,Duine2017}.
Cornelissen et al. ~\cite{Cornelissen2016} assume that at room temperature
magnon-phonon scattering of short-wavelength thermal magnons is dominated
by the exchange interaction with a scattering time of $\tau_{qp}\sim1\;\mathrm{ps}$,
which is much faster than the anisotropy-mediated magnon-phonon coupling
considered in Ref.~\cite{Rueckriegel2014} and efficiently thermalizes
magnons and phonons to equal temperatures without magnon decay. Recently,
the exchange-mediated magnon-phonon interaction~\cite{Schmidt2018}
has been taken into account in a Boltzmann approach to the SSE, but
this work underestimates the coupling strength by an order of magnitude,
as we will argue below.

In this paper we present an analytical and numerical study of magnon-phonon
interactions in bulk ferromagnetic insulators, where we take both
the anisotropy- and the exchange-mediated magnon-phonon interactions
into account. By using diagrammatic perturbation theory to calculate
the magnon self-energy, we arrive at a wave-vector dependent expression
of the magnon scattering rate, which is the inverse of the magnon
quasi-particle lifetime $\tau_{qp}$. The magnetic Grüneisen parameter
$\Gamma_{m}=\partial\ln T_{C}/\partial\ln V$ \cite{Bloch1966,Kamilov1998},
where $T_{C}$ is the Curie temperature and $V$ the volume of the
magnet, gives direct access to the exchange-mediated magnon-phonon
interaction parameter. We predict an enhancement in the phonon scattering
of the Kittel mode at the touching points of the two-magnon energy
(of the Kittel mode and a finite momentum magnon) and the longitudinal
and transverse phonon dispersions, for YIG at around $1.3\;\mathrm{T}$
and $4.6\;\mathrm{T}$. We also emphasize the difference in magnon
lifetimes that broaden light and neutron scattering experiments, and
the transport lifetimes that govern magnon heat and spin transport. 

The paper is organized as follows: in Sec.~\ref{sec:Magnons-and-phonons}
we briefly review the theory of acoustic magnons and phonons in ferro-/ferrimagnets,
particularly in YIG. In Sec.~\ref{sec:Magnon-phonon-interactions}
we derive the exchange- and anisotropy-mediated magnon-phonon interactions
for a cubic Heisenberg ferromagnet with nearest neighbor exchange
interactions in the long-wavelength limit. In Sec.~\ref{sec:Magnon-decay-rate}
we derive the magnon decay rate from the imaginary part of the magnon
self-energy in a diagrammatic approach and in Sec.~\ref{sec:Magnon-transport-lifetime}
we explain the differences between the magnon quasi-particle and transport
lifetimes. Our numerical results for YIG are discussed in Sec.~\ref{sec:Numerical-results}.
Finally in Sec.~\ref{sec:Summary} we summarize and discuss the main
results of the present work. The validity of our long-wavelength approximation
is analyzed in Appendix~\ref{sec:Longwavelength-approximation} and
in Appendix~\ref{sec:Second-order-magnetoelastic} we explain why
second order magnetoelastic couplings may be disregarded. In Appendix~\ref{sec:Numerical-integration}
we briefly discuss the numerical methods used to evaluate the k-space
integrals.

\section{Magnons and phonons in ferromagnetic insulators\label{sec:Magnons-and-phonons}}

Without loss of generality, we focus our treatment on yttrium iron
garnet (YIG). The magnon band structure of YIG has been determined
by inelastic neutron scattering \cite{Cherepanov1993,Shamoto2017,Princep2017}
and by \emph{ab initio} calculation of the exchange constants \cite{Xie2017}.
The complete magnon spectral function has been computed for all temperatures
by atomistic spin simulations \cite{Barker2016}, taking all magnon-magnon
interactions into account, but not the magnon-phonon scattering. The
pure phonon dispersion is known as well \cite{Maehrlein2016,Maehrlein2017}.
In the following, we consider the interactions of the acoustic magnons
from the lowest magnon band with transverse and longitudinal acoustic
phonons, which allows a semi-analytic treatment but limits the validity
of our results to temperatures below $100\;\mathrm{K}$. Since the
low-temperature values of the magnetoelastic constants, sound velocities,
and magnetic Grüneisen parameter are not available for YIG, we use
throughout the material parameters under ambient conditions. 

\subsection{Magnons\label{subsec:Magnons}}

Spins interact with each other via dipolar and exchange interactions.
We disregard the former since at the energy scale $E_{\mathrm{dip}}\approx0.02\;\mathrm{meV}$
\cite{Rueckriegel2014} it is only relevant for long-wavelength magnons
with wave vectors $k\lesssim6\times10^{7}\;\mathrm{m}^{-1}$ and energies
$E_{\mathbf{k}}/k_{B}\lesssim0.2\;\mathrm{K}$, which are negligible
for the thermal magnon transport in the temperature regime we are
interested in. The lowest magnon band can then be described by a simple
Heisenberg model on a course-grained simple cubic ferromagnet with
exchange interaction $J$
\begin{equation}
\mathcal{H}_{m}=-\frac{J}{2}\sum_{\left\langle i\neq j\right\rangle }\mathbf{S}_{i}\cdot\mathbf{S}_{j}-\sum_{i}g\mu_{B}BS_{i}^{z},\label{eq:Heisenberg model}
\end{equation}
where the sum is over all nearest neighbors and $\hbar\mathbf{S}_{i}$
is the spin operator at lattice site $\mathbf{R}_{i}$. The lattice
constant of the cubic lattice or YIG is $a=12.376\;\mathrm{\mathring{A}}$
and the effective spin per unit cell $\hbar S=\hbar M_{s}a^{3}/(g\mu_{B})\approx14.2\hbar$
at room temperature \cite{Rueckriegel2014} ($S\approx20$ for $T\lesssim50\;\mathrm{K}$
\cite{Maier-Flaig2017}), where the g-factor $g\approx2$, $\mu_{B}$
is the Bohr magneton and $M_{s}$ the saturation magnetization. The
parameter $J$ is an adjustable parameter that can be fitted to experiments
or computed from first principles. $B$ is an effective magnetic field
that orients the ground-state magnetization vector to the $z$ axis
and includes the (for YIG small) magnetocrystalline anisotropy field.
The $1/S$ expansion of the spin operators in terms of Holstein-Primakoff
bosons reads \cite{Holstein40},
\begin{align}
S_{i}^{+} & =S_{x}+iS_{y}\approx\sqrt{2S}\left[b_{i}+\mathcal{O}(1/S)\right],\\
S_{i}^{-} & =S_{x}-iS_{y}\approx\sqrt{2S}\left[b_{i}^{\dagger}+\mathcal{O}(1/S)\right],\\
S_{i}^{z} & =S-b_{i}^{\dagger}b_{i},
\end{align}
where $b_{i}^{\dagger}$ and $b_{i}$ are the magnon creation and
annihilation operators with boson commutation rule $\left[b_{i},b_{j}^{\dagger}\right]=\delta_{i,j}$.
Then
\begin{equation}
\mathcal{H}_{m}\rightarrow\sum_{\mathbf{k}}E_{\mathbf{k}}b_{\mathbf{k}}^{\dagger}b_{\mathbf{k}},
\end{equation}
where the magnon operators $b_{\mathbf{k}}^{\dagger}$ and $b_{\mathbf{k}}$
are defined by
\begin{align}
b_{i} & =\frac{1}{\sqrt{N}}\sum_{\mathbf{k}}e^{i\mathbf{k}\cdot\mathbf{R}_{i}}b_{\mathbf{k}},\\
b_{i}^{\dagger} & =\frac{1}{\sqrt{N}}\sum_{\mathbf{k}}e^{-i\mathbf{k}\cdot\mathbf{R}_{i}}b_{\mathbf{k}}^{\dagger},
\end{align}
and $N$ the number of unit cells. The dispersion relation 

\begin{equation}
E_{\mathbf{k}}=g\mu_{B}B+4SJ\sum_{\alpha=x,y,z}\sin^{2}(k_{\alpha}a/2)
\end{equation}
becomes quadratic in the long-wavelength limit $ka\ll1$:
\begin{equation}
E_{\mathbf{k}}=g\mu_{B}B+E_{ex}k^{2}a^{2},\label{long}
\end{equation}
where $E_{ex}=SJ$. With $E_{ex}=k_{B}\times40\;\mathrm{K}=3.45\;\mathrm{meV}$
the latter is a good approximation up to $k_{0}=1/a\approx8\times10^{8}\;\mathrm{m}^{-1}$
\cite{Cherepanov1993}. The effective exchange coupling is then $J\approx0.24\;\mathrm{meV}$.
The lowest magnon band does not depend significantly on temperature
\cite{Barker2016}, which implies that $E_{ex}=SJ$ does not depend
strongly on temperature. The temperature dependence of the saturation
magnetization and effective spin $S$ should therefore not affect
the low-energy exchange magnons significantly. By using Eq.~(\ref{long})
in the following, our theory is valid for $k\lesssim k_{0}$ (see
Fig\@.~\ref{fig:phonon dispersion}) or temperatures $T\lesssim100\;\mathrm{K}$.
In this regime the cut-off of an ultraviolet divergence does not affect
results significantly (see Appendix~\ref{sec:Longwavelength-approximation}).
We disregard magnetostatic interactions that affect the magnon spectrum
only for very small wave vectors since at low temperatures the phonon
scattering is not significant.

\subsection{Phonons}

We expand the displacement $\mathbf{X}_{i}$ of the position $\mathbf{r}_{i}$
of unit cell $i$ from the equilibrium position $\mathbf{R}_{i}$
\begin{equation}
\mathbf{X}_{i}=\mathbf{r}_{i}-\mathbf{R}_{i},
\end{equation}
into the phonon eigenmodes $X_{\mathbf{q}\lambda}$,
\begin{equation}
X_{i}^{\alpha}=\frac{1}{\sqrt{N}}\sum_{\mathbf{q},\lambda}e_{\mathbf{q}\lambda}^{\alpha}X_{\mathbf{q}\lambda}e^{i\mathbf{q}\cdot\mathbf{R}_{i}},
\end{equation}
where $\alpha\in\left\{ x,y,z\right\} $ and $\mathbf{q}$ a wave
vector. We define polarizations $\lambda\in\left\{ 1,2,3\right\} $
for the elastic continuum \cite{Flebus2017}
\begin{align}
\mathbf{e}_{\mathbf{q}1} & =\left(\cos\theta_{\mathbf{q}}\cos\phi_{\mathbf{q}},\cos\theta_{\mathbf{q}}\sin\phi_{\mathbf{q}},-\sin\theta_{\mathbf{q}}\right),\\
\mathbf{e}_{\mathbf{q}2} & =i\left(-\sin\phi_{\mathbf{q}},\cos\phi_{\mathbf{q}},0\right),\\
\mathbf{e}_{\mathbf{q}3} & =i\left(\sin\theta_{\mathbf{q}}\cos\phi_{\mathbf{q}},\sin\theta_{\mathbf{q}}\sin\phi_{\mathbf{q}},\cos\theta_{\mathbf{q}}\right),
\end{align}
where the angles $\theta_{\mathbf{q}}$ and $\phi_{\mathbf{q}}$ are
the spherical coordinates of
\begin{equation}
\mathbf{q}=q\left(\sin\theta_{\mathbf{q}}\cos\phi_{\mathbf{q}},\sin\theta_{\mathbf{q}}\sin\phi_{\mathbf{q}},\cos\theta_{\mathbf{q}}\right),
\end{equation}
which is valid for YIG up to $3\;\mathrm{THz}$ ($12\;\mathrm{meV}$)
\cite{Maehrlein2016,Maehrlein2017}. The phonon Hamiltonian then reads
\begin{align}
\mathcal{H}_{p} & =\sum_{\mathbf{q}\lambda}\left[\frac{P_{-\mathbf{q}\lambda}P_{\mathbf{q}\lambda}}{2m}+\frac{m}{2\hbar^{2}}\varepsilon_{\mathbf{q}\lambda}^{2}X_{-\mathbf{q}\lambda}X_{\mathbf{q}\lambda}\right],\nonumber \\
 & =\sum_{\mathbf{q}\lambda}\varepsilon_{\mathbf{q}\lambda}\left(a_{\mathbf{q}\lambda}^{\dagger}a_{\mathbf{q}\lambda}+\frac{1}{2}\right),
\end{align}
where the canonical momenta $P_{\mathbf{q}\lambda}$ obey the commutation
relations $\left[X_{\mathbf{q}\lambda},P_{\mathbf{q}'\lambda'}\right]=i\hbar\delta_{\mathbf{q},-\mathbf{q}'}\delta_{\lambda\lambda'}$
and the mass of the YIG unit cell $m=\rho a^{3}=9.8\times10^{-24}\;\mathrm{kg}$
\cite{Gurevich1996}. The phonon dispersions for YIG then read
\begin{equation}
\varepsilon_{\mathbf{q}\lambda}=\hbar c_{\lambda}|\mathbf{q}|,
\end{equation}
where $c_{1,2}=c_{t}=3843\;\mathrm{m/s}$ is the transverse sound
velocity and $c_{3}=c_{l}=7209\;\mathrm{m/s}$ the longitudinal velocity
at room temperature \cite{Gurevich1996}. In terms of phonon creation
and annihilation operators
\begin{equation}
X_{\mathbf{q}\lambda}=\frac{a_{\mathbf{q\lambda}}+a_{-\mathbf{q}\lambda}^{\dagger}}{\sqrt{2m\varepsilon_{\mathbf{q}\lambda}/\hbar^{2}}},\quad P_{\mathbf{q}\lambda}=\frac{1}{i}\sqrt{\frac{m\varepsilon_{\mathbf{q}\lambda}}{2}}\left(a_{\mathbf{q\lambda}}-a_{-\mathbf{q}\lambda}^{\dagger}\right),
\end{equation}
and $\left[a_{\mathbf{q}\lambda},a_{\mathbf{q}'\lambda'}^{\dagger}\right]=\delta_{\mathbf{q},\mathbf{q}'}\delta_{\lambda,\lambda'}$.

\begin{figure}
\begin{centering}
\includegraphics{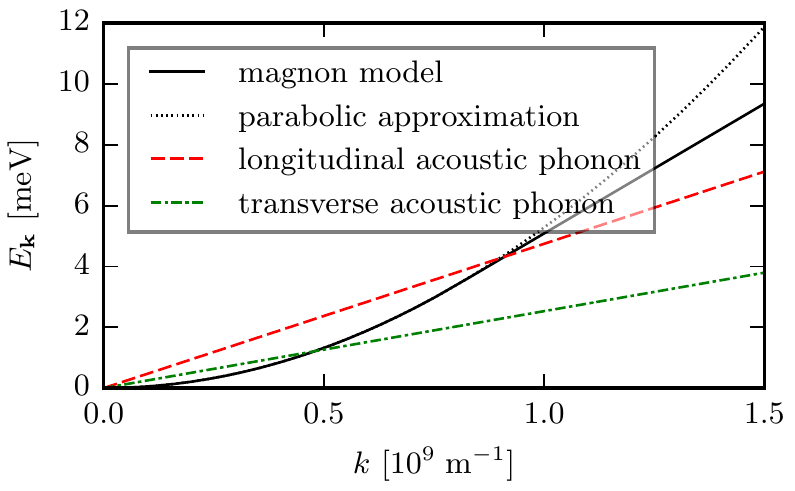}
\par\end{centering}
\caption{Dispersion relations of the acoustic phonons and magnons in YIG at
zero magnetic field. \label{fig:phonon dispersion}}
\end{figure}

In Fig.~\ref{fig:phonon dispersion} we plot the longitudinal and
transverse phonon and the acoustic magnon dispersion relations for
YIG at zero magnetic field. The magnon-phonon interaction leads to
an avoided level crossing at points where magnon and phonon dispersion
cross, as discussed in Refs.~\cite{Gurevich1996} and \cite{Rueckriegel2014}.

\section{Magnon-phonon interactions\label{sec:Magnon-phonon-interactions}}

We derive in this section the magnon-phonon interactions due to the
anisotropy and exchange interactions for a cubic lattice ferromagnet.

\subsection{Phenomenological magnon-phonon interaction}

In the long-wavelength/continuum limit ($k\lesssim k_{0}$) the magnetoelastic
energy to lowest order in the deviations of magnetization and lattice
from equilibrium reads \cite{Rueckriegel2014,Abrahams1952,Kittel1953,Kittel1958,Kaganov1959}

\begin{align}
E_{me}= & \frac{n}{M_{s}^{2}}\int d^{3}r\sum_{\alpha\beta}\left[B_{\alpha\beta}M_{\alpha}(\mathbf{r})M_{\beta}(\mathbf{r})\right.\nonumber \\
 & \left.+B'_{\alpha\beta}\frac{\partial\mathbf{M}(\mathbf{r})}{\partial r_{\alpha}}\cdot\frac{\partial\mathbf{M}(\mathbf{r})}{\partial r_{\beta}}\right]X_{\alpha\beta}(\mathbf{r}),\label{eq:magnetoelastic energy}
\end{align}
where $n=1/a^{3}$. The strain tensor $X_{\alpha\beta}$ is defined
in terms of the lattice displacements $X_{\alpha}$,
\begin{equation}
X_{\alpha\beta}(\mathbf{r})=\frac{1}{2}\left[\frac{\partial X_{\alpha}(\mathbf{r})}{\partial r_{\beta}}+\frac{\partial X_{\beta}(\mathbf{r})}{\partial r_{\alpha}}\right],
\end{equation}
with, for a cubic lattice \cite{Rueckriegel2014},
\begin{align}
B_{\alpha\beta} & =\delta_{\alpha\beta}B_{\parallel}+(1-\delta_{\alpha\beta})B_{\perp},\\
B'_{\alpha\beta} & =\delta_{\alpha\beta}B'_{\parallel}+(1-\delta_{\alpha\beta})B'_{\perp}.
\end{align}
$B_{\alpha\beta}$ is caused by magnetic anisotropies and $B_{\alpha\beta}'$
by the exchange interaction under lattice deformations. For YIG at
room temperature \cite{Gurevich1996,Kamilov1998}
\begin{align}
B_{\parallel} & =k_{B}\times47.8\;\mathrm{K}=4.12\;\mathrm{meV},\\
B_{\perp} & =k_{B}\times95.6\;\mathrm{K}=8.24\;\mathrm{meV},\\
B_{\parallel}'/a^{2} & =k_{B}\times2727\;\mathrm{K}=235\;\mathrm{meV},\\
B_{\perp}'/a^{2} & \approx0.
\end{align}
We discuss the values for $B_{\parallel}'$ and $B'_{\perp}$ in Sec.~\ref{subsec:Exchange-mediated}.

\subsection{Anisotropy-mediated magnon-phonon interaction}

The magnetoelastic anisotropy (\ref{eq:magnetoelastic energy}) is
described by the Hamiltonian \cite{Rueckriegel2014},

\begin{align}
\mathcal{H}_{mp}^{an} & =\sum_{\mathbf{q}\lambda}\left[\Gamma_{\mathbf{q}\lambda}b_{-\mathbf{q}}X_{\mathbf{q}\lambda}+\Gamma_{-\mathbf{q}\lambda}^{*}b_{\mathbf{q}}^{\dagger}X_{\mathbf{q}\lambda}\right]\nonumber \\
 & +\frac{1}{\sqrt{N}}\sum_{\mathbf{q},\mathbf{k},\mathbf{k}'}\delta_{\mathbf{k}-\mathbf{k}'-\mathbf{q},0}\sum_{\lambda}\Gamma_{\mathbf{kk}',\lambda}^{an}b_{\mathbf{k}}^{\dagger}b_{\mathbf{k}'}X_{\mathbf{q}\lambda}\nonumber \\
 & +\frac{1}{\sqrt{N}}\sum_{\mathbf{q},\mathbf{k},\mathbf{k}'}\delta_{\mathbf{k}+\mathbf{k}'+\mathbf{q},0}\sum_{\lambda}\Gamma_{\mathbf{kk}',\lambda}^{bb}b_{\mathbf{k}}b_{\mathbf{k}'}X_{\mathbf{q}\lambda}\nonumber \\
 & +\frac{1}{\sqrt{N}}\sum_{\mathbf{q},\mathbf{k},\mathbf{k}'}\delta_{\mathbf{k}+\mathbf{k}'-\mathbf{q},0}\sum_{\lambda}\Gamma_{\mathbf{kk}',\lambda}^{\bar{b}\bar{b}}b_{\mathbf{k}}^{\dagger}b_{\mathbf{k}'}^{\dagger}X_{\mathbf{q}\lambda},\label{eq:an magnon-phonon}
\end{align}
with interaction vertices

\begin{align}
\Gamma_{\mathbf{q}\lambda} & =\frac{B_{\perp}}{\sqrt{2S}}\left[iq_{z}e_{\mathbf{q}\lambda}^{x}+q_{z}e_{\mathbf{q}\lambda}^{y}\right.\nonumber \\
 & \left.+\left(iq_{x}+q_{y}\right)e_{\mathbf{q}\lambda}^{z}\right],\\
\Gamma_{\mathbf{kk}',\lambda}^{an} & =U_{\mathbf{k}-\mathbf{k}',\lambda},\\
\Gamma_{\mathbf{kk}',\lambda}^{bb} & =V_{-\mathbf{k}-\mathbf{k}',\lambda},\label{eq:Gamma^bb}\\
\Gamma_{\mathbf{kk}',\lambda}^{\bar{b}\bar{b}} & =V_{-\mathbf{k}-\mathbf{k}',\lambda}^{*},\label{eq:Gamma^b*b*}
\end{align}
and 
\begin{align}
U_{\mathbf{q},\lambda} & =\frac{iB_{\parallel}}{S}\left[q_{x}e_{\mathbf{q}\lambda}^{x}+q_{y}e_{\mathbf{q}\lambda}^{y}-2q_{z}e_{\mathbf{q}\lambda}^{z}\right],\\
V_{\mathbf{q},\lambda} & =\frac{iB_{\parallel}}{S}\left[q_{x}e_{\mathbf{q}\lambda}^{x}-q_{y}e_{\mathbf{q}\lambda}^{y}\right]\nonumber \\
 & +\frac{B_{\perp}}{S}\left[q_{y}e_{\mathbf{q}\lambda}^{x}+q_{x}e_{\mathbf{q}\lambda}^{y}\right].\label{eq:V}
\end{align}

The one magnon-two phonon process is of the same order in the total
number of magnons and phonons as the two magnon-one phonon processes,
but its effect on magnon transport is small, as shown in Appendix~\ref{sec:Second-order-magnetoelastic}.

\subsection{Exchange-mediated magnon-phonon interaction\label{subsec:Exchange-mediated}}

The exchange-mediated magnon-phonon interaction is obtained under
the assumption that the exchange interaction $J_{ij}$ between two
neighboring spins at lattice sites $\mathbf{r}_{i}$ and $\mathbf{r}_{j}$
depends only on their distance, which leads to the expansion to leading
order in the small parameter $(|\mathbf{r}_{i}-\mathbf{r}_{j}|-a)$
\begin{equation}
J_{ij}=J(|\mathbf{r}_{i}-\mathbf{r}_{j}|)\approx J+J'\cdot(|\mathbf{r}_{i}-\mathbf{r}_{j}|-a),\label{eq:J expansion}
\end{equation}
where $a$ is the equilibrium distance and $J'=\partial J/\partial a$.
With $\mathbf{r}_{i}=\mathbf{R}_{i}+\mathbf{X}_{\mathbf{R}_{i}},$
the Heisenberg Hamiltonian (\ref{eq:Heisenberg model}) is modulated
by
\begin{equation}
\mathcal{H}_{mp}^{ex}=-J'\sum_{i}\sum_{\alpha=x,y,z}\left(X_{\mathbf{R}_{i}+a\mathbf{e}_{\alpha}}^{\alpha}-X_{\mathbf{R}_{i}}^{\alpha}\right)\mathbf{S}_{\mathbf{R}_{i}}\cdot\mathbf{S}_{\mathbf{R}_{i}+a\mathbf{e}_{\alpha}},\label{eq:exchange magnon-phonon}
\end{equation}
where $\mathbf{e}_{\alpha}$ is a unit vectors in the $\alpha$ direction.
Expanding the displacements in terms of the phonon and magnon modes
\begin{equation}
\mathcal{H}_{mp}^{ex}=\frac{1}{\sqrt{N}}\sum_{\mathbf{q},\mathbf{k},\mathbf{k}'}\delta_{\mathbf{k}-\mathbf{k}'-\mathbf{q},0}\sum_{\lambda}\Gamma_{\mathbf{kk}',\lambda}^{ex}b_{\mathbf{k}}^{\dagger}b_{\mathbf{k}'}X_{\mathbf{q}\lambda},\label{eq:exchange magnon-phonon-2}
\end{equation}
with interaction
\begin{align}
\Gamma_{\mathbf{kk}',\lambda}^{ex} & =8iJ'S\sum_{\alpha}e_{\mathbf{k}-\mathbf{k}',\lambda}^{\alpha}\sin\left(\frac{k_{\alpha}a}{2}\right)\sin\left(\frac{k'_{\alpha}a}{2}\right)\nonumber \\
 & \times\sin\left(\frac{(k_{\alpha}-k'_{\alpha})a}{2}\right)\nonumber \\
 & \approx iJ'a^{3}S\sum_{\alpha}e_{\mathbf{k}-\mathbf{k}',\lambda}^{\alpha}k_{\alpha}k_{\alpha'}\left(k_{\alpha}-k'_{\alpha}\right),
\end{align}
where the last line is the long-wavelength expansion. The magnon-phonon
interaction 
\begin{equation}
\Gamma_{\mathbf{k},\mathbf{k}',\lambda}^{\bar{b}b}=\Gamma_{\mathbf{k},\mathbf{k}',\lambda}^{ex}+\Gamma_{\mathbf{k},\mathbf{k}',\lambda}^{an}\label{eq:Gamma_c}
\end{equation}
conserves the magnon number, while (\ref{eq:Gamma^bb}) and (\ref{eq:Gamma^b*b*})
do not. Phonon numbers are not conserved in either case.

The value of $J'$ for YIG is determined by the magnetic Grüneisen
parameter \cite{Bloch1966,Kamilov1998}
\begin{equation}
\Gamma_{m}=\frac{\partial\ln T_{C}}{\partial\ln V}=\frac{\partial\ln J}{\partial\ln V}=\frac{J'a}{3J},
\end{equation}
where $V=Na^{3}$ is the volume of the magnet. The only assumption
here is that the Curie temperature $T_{C}$ scales linearly with the
exchange constant $J$ \cite{Blundell2001}. $\Gamma_{m}$ has been
measured for YIG via the compressibility to be $\Gamma_{m}=-3.26$
\cite{Bloch1966}, and via thermal expansion, $\Gamma_{m}=-3.13$
\cite{Kamilov1998}, so we set $\Gamma_{m}=-3.2$. For other materials
the magnetic Grüneisen parameter is also of the order of unity and
in many cases $\Gamma_{m}\approx-10/3$ \cite{Bloch1966,Samara1969,Kamilov1998}.
A recent \emph{ab initio} study of YIG finds $\Gamma_{m}=-3.1$ \cite{Liu2017}. 

Comparing the continuum limit of Eq. (\ref{eq:exchange magnon-phonon})
with the classical magnetoelastic energy (\ref{eq:magnetoelastic energy})
\begin{equation}
B_{\parallel}'=3\Gamma_{m}JS^{2}a^{2}/2,
\end{equation}
where for YIG $B_{\parallel}'/a^{2}\approx235\;\mathrm{meV}$. We
disregard $B_{\perp}'$ since it vanishes for nearest neighbor interactions
by cubic lattice symmetry.

The coupling strength of the exchange-mediated magnon-phonon interaction
can be estimated from the exchange energy $SJ'a\approx E_{ex}=SJ$
\cite{Shklovskij2018,Schmidt2018} following Akhiezer \emph{et al.}
\cite{Akhiezer1961a,Akhiezer1961b}. Our estimate of $SJ'a=3\Gamma_{m}SJ$
is larger by $3\Gamma_{m}$, i.e. one order of magnitude. Since the
scattering rate is proportional to the square of the interaction strength,
our estimate of the scattering rate is a factor $100$ larger than
previous ones. The assumption $J'a\approx J$ is too small to be consistent
with the experimental Grüneisen constant \cite{Bloch1966,Kamilov1998}.
Ref.~\cite{Cornelissen2016} educatedly guessed $J'a\approx100J,$
which we now judge to be too large.

\subsection{Interaction vertices}

\begin{figure}[t]
\begin{centering}
\includegraphics[width=3.17622in]{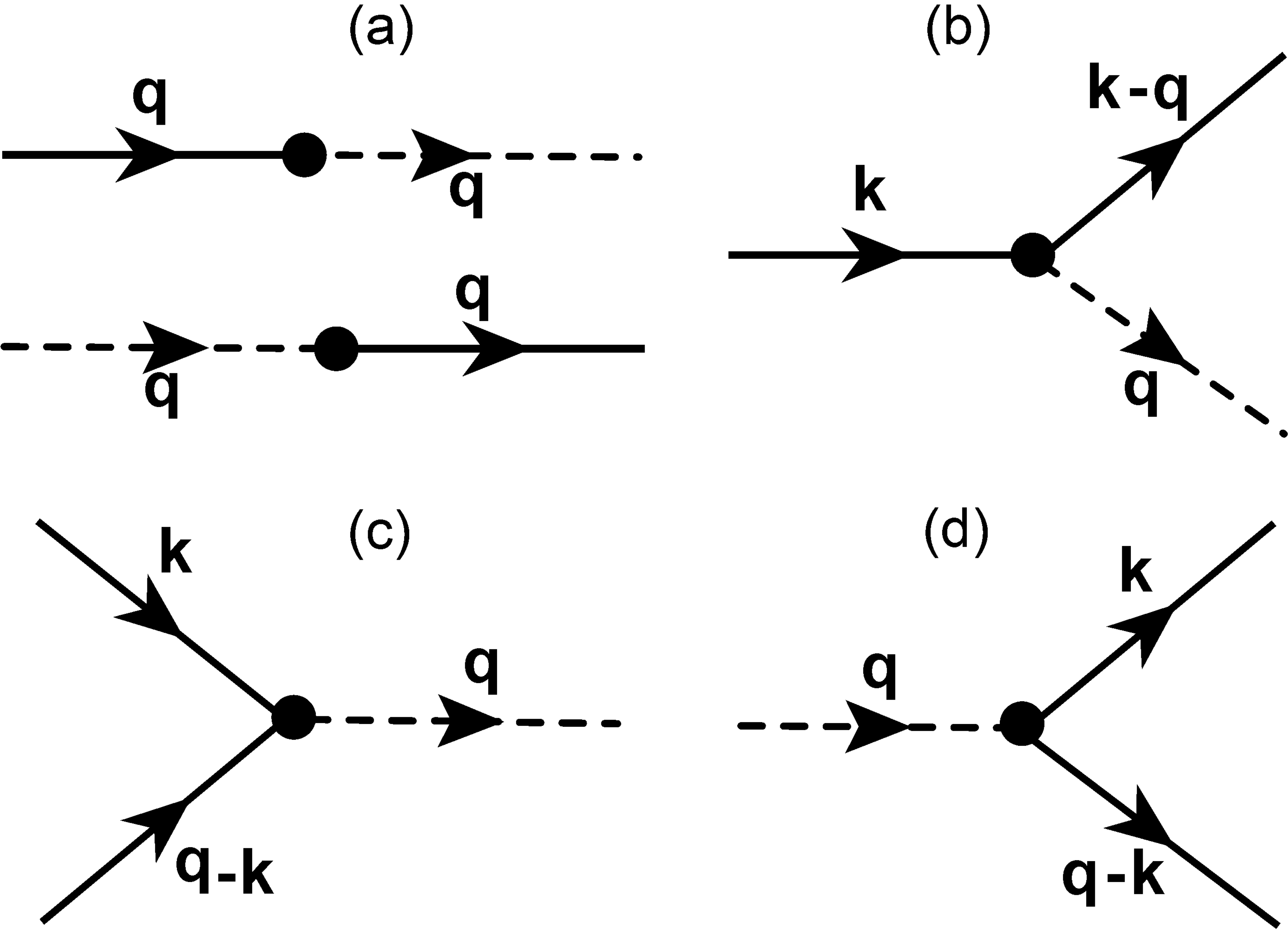}
\par\end{centering}
\caption{Feynman diagrams of interactions between magnons (solid lines) and
phonons (dashed lines). The arrows indicate the energy-momentum flow.
(a) magnon-phonon interconversion, (b) magnon number-conserving magnon-phonon
interaction, (c) and (d) magnon number non-conserving magnon-phonon
interactions.\label{fig:interactions}}
\end{figure}
The magnon-phonon interactions in the Hamiltonian (\ref{eq:an magnon-phonon})
are shown in Fig.~\ref{fig:interactions} as Feynman diagrams. Fig.~\ref{fig:interactions}(a)
illustrates magnon and phonon interconversion, which is responsible
for the magnon-phonon hybridization and level splitting at the crossing
of magnon and phonon dispersions \cite{Gurevich1996,Rueckriegel2014}.
The divergence of this diagram at the magnon-phonon crossing points
is avoided by either direct diagonalization of the magnon-phonon Hamiltonian
\cite{Flebus2017} or by cutting-off the divergence by a lifetime
parameter \cite{Schmidt2018}. This process still generates enhanced
magnon transport that is observable as magnon polaron anomalies in
the spin Seebeck effect \cite{Kikkawa2016} or spin-wave excitation
thresholds \cite{Turner1960,Gurevich1975}, but these are strongly
localized in phase space and disregarded in the following, where we
focus on the magnon scattering rates to leading order in $1/S$ of
the scattering processes in Fig.~\ref{fig:interactions}(b)-(d). 

\section{Magnon scattering rate\label{sec:Magnon-decay-rate}}

Here we derive the magnon reciprocal quasi-particle lifetime $\tau_{qp}^{-1}=\gamma$
as the imaginary part of the wave vector dependent self-energy, caused
by acoustic phonon scattering \cite{Rueckriegel2014},
\begin{equation}
\gamma(\mathbf{k})=-\frac{2}{\hbar}\mathrm{Im}\Sigma(\mathbf{k},E_{\mathbf{k}}/\hbar+i0^{+}).\label{eq:decay rate from self-energy}
\end{equation}
This quantity is in principle observable by inelastic neutron scattering.
The total decay rate 
\begin{equation}
\gamma=\gamma^{c}+\gamma^{nc}+\gamma^{\mathrm{other}}
\end{equation}
is the sum of the magnon number conserving decay rate $\gamma^{c}$
and the magnon number non-conserving decay rate $\gamma^{nc}$, which
are related to the magnon-phonon scattering time $\tau_{mp}$ and
the magnon-phonon dissipation time $\tau_{mr}$ by
\begin{equation}
\tau_{mp}=\frac{1}{\gamma^{c}},\quad\tau_{mr}=\frac{1}{\gamma^{nc}}.
\end{equation}
$\gamma^{\mathrm{other}}$ is caused by magnon-magnon and magnon disorder
scattering, thereby beyond the scope of this work.

\begin{figure}
\begin{centering}
\includegraphics[width=3.17622in]{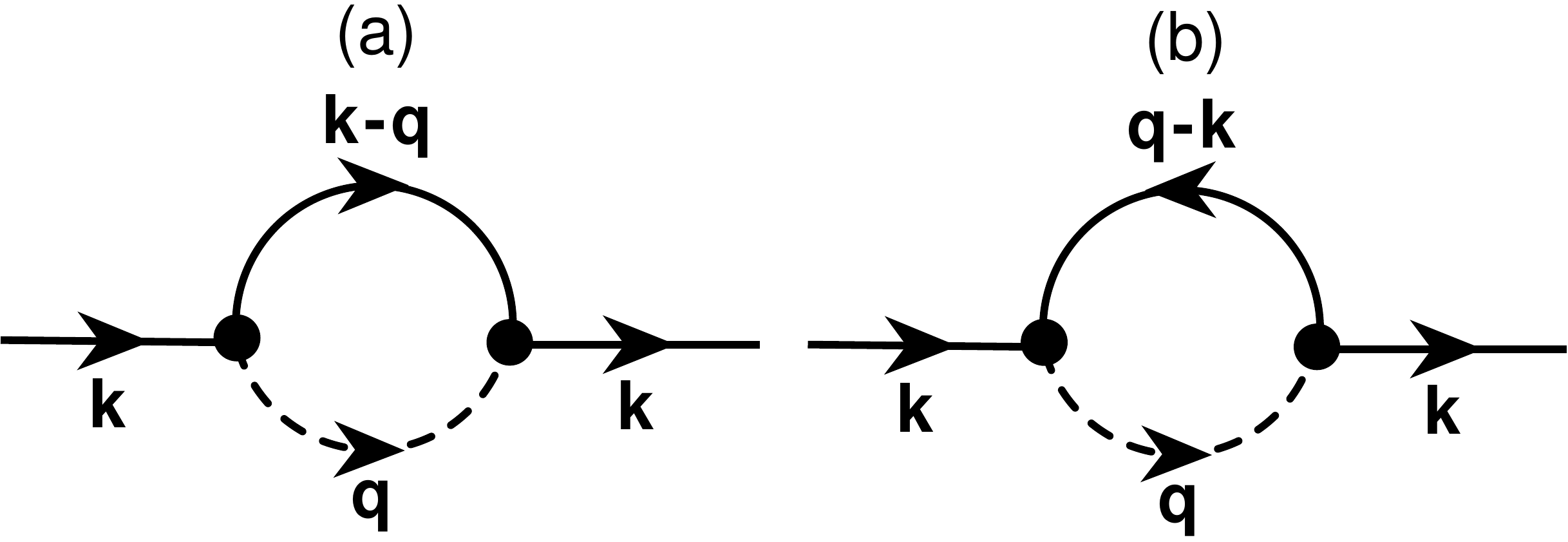}
\par\end{centering}
\caption{Feynman diagrams representing the self-energy Eq.~(\ref{eq:self-energy})
due to (a) magnon number-conserving magnon-phonon interactions and
(b) magnon number non-conserving magnon-phonon interactions.\label{fig:self-energy}}
\end{figure}

The self-energy to leading order in the $1/S$ expansion is of second
order in the magnon-phonon interaction \cite{Rueckriegel2014},\begin{widetext}

\begin{align}
\Sigma_{2}(\mathbf{k},i\omega) & =\frac{1}{N}\sum_{\mathbf{k'}\lambda}\frac{\hbar^{2}\left|\Gamma_{\mathbf{k},\mathbf{k}',\lambda}^{\bar{b}b}\right|^{2}}{2m\varepsilon_{\mathbf{k}-\mathbf{k}',\lambda}}\left[\frac{n_{B}(\varepsilon_{\mathbf{k}-\mathbf{k}',\lambda})-n_{B}(E_{\mathbf{k}'})}{i\hbar\omega+\varepsilon_{\mathbf{k}-\mathbf{k}',\lambda}-E_{\mathbf{k}'}}+\frac{1+n_{B}(\varepsilon_{\mathbf{k}-\mathbf{k}',\lambda})+n_{B}(E_{\mathbf{k}'})}{i\hbar\omega-\varepsilon_{\mathbf{k}-\mathbf{k}',\lambda}-E_{\mathbf{k}'}}\right]\nonumber \\
 & -\frac{1}{N}\sum_{\mathbf{k'}\lambda}\frac{\hbar^{2}\left|\Gamma_{\mathbf{k},\mathbf{k}',\lambda}^{bb}\right|^{2}}{2m\varepsilon_{\mathbf{k}-\mathbf{k}',\lambda}}\left[\frac{1+n_{B}(\varepsilon_{\mathbf{k}+\mathbf{k}',\lambda})+n_{B}(E_{\mathbf{k}'})}{i\hbar\omega+\varepsilon_{\mathbf{k}+\mathbf{k}',\lambda}+E_{\mathbf{k}'}}+\frac{n_{B}(\varepsilon_{\mathbf{k}+\mathbf{k}',\lambda})-n_{B}(E_{\mathbf{k}'})}{i\hbar\omega-\varepsilon_{\mathbf{k}+\mathbf{k}',\lambda}+E_{\mathbf{k}'}}\right],\label{eq:self-energy}
\end{align}
\end{widetext}where the magnon number conserving magnon-phonon scattering
vertex $\Gamma_{\mathbf{k},\mathbf{k}',\lambda}^{\bar{b}b}=\Gamma_{\mathbf{k},\mathbf{k}',\lambda}^{ex}+\Gamma_{\mathbf{k},\mathbf{k}',\lambda}^{an}$
and the Planck (Bose) distribution function $n_{B}(\varepsilon)=(e^{\beta\varepsilon}-1)^{-1}$
with inverse temperature $\beta=1/(k_{B}T)$. The Feynman diagrams
representing the magnon number conserving and non-conserving contributions
to the self-energy are shown in Fig.~\ref{fig:self-energy}.

We write the decay rate in terms of four contributions
\begin{equation}
\gamma(\mathbf{k})=\gamma_{out}^{c}(\mathbf{k})+\gamma_{out}^{nc}(\mathbf{k})-\gamma_{in}^{c}(\mathbf{k})-\gamma_{in}^{nc}(\mathbf{k}),
\end{equation}
where $out$ and $in$ denote the out-scattering and in-scattering
parts. The contributions to the decay rate read \cite{Rueckriegel2014}

\begin{widetext}

\begin{align}
\gamma_{out}^{c}(\mathbf{k})= & \frac{\pi\hbar}{mN}\sum_{\mathbf{q},\lambda}\frac{\left|\Gamma_{\mathbf{k},\mathbf{k}-\mathbf{q},\lambda}^{\bar{b}b}\right|^{2}}{\varepsilon_{\mathbf{q}\lambda}}\left[(1+n_{B}(E_{\mathbf{k}-\mathbf{q}}))n_{B}(\varepsilon_{\mathbf{q}\lambda})\delta(E_{\mathbf{k}}-E_{\mathbf{k}-\mathbf{q}}+\varepsilon_{\mathbf{q}\lambda})\right.\nonumber \\
 & \phantom{{+\sum_{\mathbf{q},\lambda}\frac{\left|\Gamma_{\mathbf{k},\mathbf{k}-\mathbf{q},\lambda}^{\bar{b}b}\right|^{2}}{\varepsilon_{\mathbf{q}\lambda}}\}}}+\left.(1+n_{B}(E_{\mathbf{k}-\mathbf{q}}))(1+n_{B}(\varepsilon_{\mathbf{q}\lambda}))\delta(E_{\mathbf{k}}-E_{\mathbf{k}-\mathbf{q}}-\varepsilon_{\mathbf{q}\lambda})\right],\\
\gamma_{in}^{c}(\mathbf{k})= & \frac{\pi\hbar}{mN}\sum_{\mathbf{q},\lambda}\frac{\left|\Gamma_{\mathbf{k},\mathbf{k}-\mathbf{q},\lambda}^{\bar{b}b}\right|^{2}}{\varepsilon_{\mathbf{q}\lambda}}\left[n_{B}(E_{\mathbf{k}-\mathbf{q}})(1+n_{B}(\varepsilon_{\mathbf{q}\lambda}))\delta(E_{\mathbf{k}}-E_{\mathbf{k}-\mathbf{q}}+\varepsilon_{\mathbf{q}\lambda})\right.\nonumber \\
 & \phantom{{+\sum_{\mathbf{q},\lambda}\frac{\left|\Gamma_{\mathbf{k},\mathbf{k}-\mathbf{q},\lambda}^{\bar{b}b}\right|^{2}}{\varepsilon_{\mathbf{q}\lambda}}\}}}+\left.n_{B}(E_{\mathbf{k}-\mathbf{q}})n_{B}(\varepsilon_{\mathbf{q}\lambda})\delta(E_{\mathbf{k}}-E_{\mathbf{k}-\mathbf{q}}-\varepsilon_{\mathbf{q}\lambda})\right],\\
\gamma_{out}^{nc}(\mathbf{k})= & \frac{\pi\hbar}{mN}\sum_{\mathbf{q},\lambda}\frac{\left|\Gamma_{\mathbf{k},\mathbf{q}-\mathbf{k},\lambda}^{bb}\right|^{2}}{\varepsilon_{\mathbf{q}\lambda}}\left[n_{B}(E_{\mathbf{q}-\mathbf{k}})(1+n_{B}(\varepsilon_{\mathbf{q}\lambda}))\delta(E_{\mathbf{k}}+E_{\mathbf{q}-\mathbf{k}}-\varepsilon_{\mathbf{q}\lambda})\right],\label{eq:gamma_nc_out}\\
\gamma_{in}^{nc}(\mathbf{k})= & \frac{\pi\hbar}{mN}\sum_{\mathbf{q},\lambda}\frac{\left|\Gamma_{\mathbf{k},\mathbf{q}-\mathbf{k},\lambda}^{bb}\right|^{2}}{\varepsilon_{\mathbf{q}\lambda}}\left[(1+n_{B}(E_{\mathbf{q}-\mathbf{k}}))n_{B}(\varepsilon_{\mathbf{q}\lambda})\delta(E_{\mathbf{k}}+E_{\mathbf{q}-\mathbf{k}}-\varepsilon_{\mathbf{q}\lambda})\right],\label{eq:gamma_nc_in}
\end{align}

\end{widetext}where the sum is over all momenta $\mathbf{q}$ in
the Brillouin zone. Here the magnon/phonon annihilation rate is proportional
to the Boson number $n_{B}$, while the creation rate scales with
$1+n_{B}$. For example, in the out-scattering rate $\gamma_{out}^{c}(\mathbf{k})$
the incoming magnon with momentum $\mathbf{k}$ gets scattered into
the state $\mathbf{k}-\mathbf{q}$ and a phonon is either absorbed
with probability $\sim n_{B}$ or emitted with probability $\sim(1+n_{B})$.
The out- and in-scattering rates are related by the detailed balance
\begin{equation}
\gamma_{in}^{c}(\mathbf{k})/\gamma_{out}^{c}(\mathbf{k})=\gamma_{in}^{nc}(\mathbf{k})/\gamma_{out}^{nc}(\mathbf{k})=e^{-\beta E_{\mathbf{k}}}.
\end{equation}
For high temperatures $k_{B}T\gg E_{\mathbf{k}}$, we may expand the
Bose functions $n_{B}(E_{\mathbf{k}})\sim k_{B}T/E_{\mathbf{k}}$
and we find $\gamma_{in}\sim\gamma_{out}\sim T^{2}$ and $\gamma=\gamma_{out}-\gamma_{in}\sim T$.
For low temperatures $k_{B}T\ll E_{\mathbf{k}}$, the out-scattering
rate $\gamma_{out}\to\mathrm{const}.$ and the in-scattering rate
$\gamma_{in}\sim e^{-\beta E_{\mathbf{k}}}\to0$. The scattering processes
(c) and (d) in Fig.~\ref{fig:interactions} conserve energy and linear
momentum, but not angular momentum. A loss of angular momentum after
integration over all wave vectors corresponds to a mechanical torque
on the total lattice that contributes to the Einstein-de Haas effect
\cite{Nakane2018}.

\section{Magnon transport lifetime\label{sec:Magnon-transport-lifetime}}

In this section we compare the transport lifetime $\tau_{t}$ and
the magnon quasi-particle lifetime $\tau_{qp}$ that can be very different
\cite{Mahan,DasSarma1985,Hwang2008}, but, to the best of our knowledge,
has not yet been addressed for magnons. The magnon decay rate is proportional
to the imaginary part of self energy, as shown in Eq.~(\ref{eq:decay rate from self-energy}).
On the other hand, the transport is governed by transport lifetime
$\tau_{t}$ in the Boltzmann equation that agrees with $\tau_{qp}$
only in the relaxation time approximation. The stationary Boltzmann
equation for the magnon distribution can be written as \cite{Cornelissen2016,Flebus2017}
\begin{equation}
\frac{\partial f_{\mathbf{k}}(\mathbf{r})}{\partial\mathbf{r}}\cdot\frac{\partial E_{\mathbf{k}}}{\partial(\hbar\mathbf{k})}=\Gamma_{in}[f]-\Gamma_{out}[f],\label{eq:Boltzmann equation}
\end{equation}
where $f_{\mathbf{k}}(\mathbf{r})$ is the magnon distribution function.
The $in$ and $out$ contributions to the collision integral are related
to the previously defined in- and out-scattering rates by 
\begin{align}
\Gamma_{in}[f] & =(1+f_{\mathbf{k}})\gamma_{in}[f],\\
\Gamma_{out}[f] & =f_{\mathbf{k}}\gamma_{out}[f],
\end{align}
where the equilibrium magnon distribution $n_{B}(E_{\mathbf{k}})$
is replaced by the non-equilibrium distribution function $f_{\mathbf{k}}$.
The factor $(1+f_{\mathbf{k}})$ corresponds to the creation of a
magnon with momentum $\mathbf{k}$ in the in-scattering process and
the factor $f_{\mathbf{k}}$ to the annihilation in the out-scattering
process. The phonons are assumed to remain at thermal equilibrium,
so we disregard the phonon drift contribution that is expected in
the presence of a phononic heat current.

Magnon transport is governed by three linear response functions, i.e.
spin and heat conductivity and spin Seebeck coefficient \cite{Flebus2017}.
These can be obtained from the expansion of the distribution function
in terms of temperature and chemical potential gradients and correspond
to two-particle Green functions with vertex corrections, that reflect
the non-equilibrium in-scattering processes, captured by a transport
lifetime $\tau_{t}$ that can be different from the quasi-particle
(dephasing) lifetime $\tau_{qp}$ defined by the self-energy. We define
the transport life time of a magnon with momentum $\mathbf{k}$ in
terms of the collision integral 
\begin{equation}
\Gamma_{out}[f]-\Gamma_{in}[f]=\frac{1}{\tau_{\mathbf{k},t}[f]}\left(f_{\mathbf{k}}(\mathbf{r})-f_{0,\mathbf{k}}\right),\label{eq:relaxation time approximation}
\end{equation}
with $f_{0,\mathbf{k}}=n_{B}(E_{\mathbf{k}})$ and we assume a thermalized
quasi-equilibrium distribution function 
\begin{equation}
f_{\mathbf{k}}(\mathbf{r})=n_{B}\left(\frac{E_{\mathbf{k}}-\mu(\mathbf{r})}{k_{B}T(\mathbf{r})}\right),
\end{equation}
where $\mu$ is the magnon chemical potential. We linearize the function
$f_{\mathbf{k}}$ in terms of small deviations $\delta f_{\mathbf{k}}$
from equilibrium $f_{0,\mathbf{k}}$, 
\begin{equation}
\delta f_{\mathbf{k}}=f_{\mathbf{k}}-f_{0,\mathbf{k}}.\label{eq:f}
\end{equation}
leading to \cite{Cornelissen2016}
\begin{equation}
\delta f_{\mathbf{k}}=\tau_{\mathbf{k},t}\left[f\right]\frac{\partial f_{0,\mathbf{k}}}{\partial E_{\mathbf{k}}}\frac{\partial E_{\mathbf{k}}}{\partial(\hbar\mathbf{k})}\cdot\left(\boldsymbol{\nabla}\mu+\frac{E_{\mathbf{k}}-\mu}{T}\boldsymbol{\nabla}T\right),\label{eq:magnon shift}
\end{equation}
where the gradients of chemical potential $\boldsymbol{\nabla}\mu$
and temperature $\boldsymbol{\nabla}T$ drive the magnon current.
In the relaxation time approximation we disregard the dependence of
$\tau_{\mathbf{k},t}[f]$ on $\delta f$ and recover the quasi-particle
lifetime $\tau_{\mathbf{k},t}\rightarrow\tau_{\mathbf{k},qp}$.

To first order in the phonon operators and second order in the magnon
operators the collision integral for magnon number non-conserving
processes, 
\begin{align}
\Gamma_{out}^{nc}[f]-\Gamma_{in}^{nc}[f]\nonumber \\
 & \hspace{-2cm}=\frac{\pi\hbar}{mN}\sum_{\mathbf{q}\lambda}\frac{\vert\Gamma_{\mathbf{k},\mathbf{\mathbf{q}-\mathbf{k}},\lambda}^{bb}\vert^{2}}{\varepsilon_{\mathbf{q}\lambda}}\delta(E_{\mathbf{k}}+E_{\mathbf{\mathbf{q}-\mathbf{k}}}-\varepsilon_{\mathbf{q}\lambda})\nonumber \\
 & \hspace{-2cm}\times\left[(1+n_{\mathbf{q}\lambda})f_{\mathbf{k}}f_{\mathbf{q}-\mathbf{k}}-n_{\mathbf{\mathbf{q}\lambda}}(1+f_{\mathbf{q}-\mathbf{k}})(1+f_{\mathbf{k}})\right],\label{eq:collision integral nc}
\end{align}
where the interaction vertex $\Gamma_{\mathbf{k},\mathbf{k'},\lambda}^{bb}$
is given by Eq.~(\ref{eq:Gamma^bb}) and $n_{\mathbf{\mathbf{q}\lambda}}=n_{B}(\varepsilon_{\mathbf{\mathbf{q}\lambda}})$.
By using the expansion (\ref{eq:f}) in the collision integral that
vanishes at equilibrium,
\begin{equation}
\Gamma_{out}[f_{0}]-\Gamma_{in}[f_{0}]=0,
\end{equation}
we arrive at

\begin{align}
\frac{1}{\tau_{\mathbf{k},t}^{nc}} & =\frac{\pi\hbar}{mN}\sum_{\mathbf{q}\lambda}\frac{\vert\Gamma_{\mathbf{k},\mathbf{\mathbf{q}-\mathbf{k}},\lambda}^{bb}\vert^{2}}{\varepsilon_{\mathbf{q}\lambda}}\delta(E_{\mathbf{k}}+E_{\mathbf{\mathbf{q}-\mathbf{k}}}-\varepsilon_{\mathbf{q}\lambda})\nonumber \\
 & \times\left[n_{B}(E_{\mathbf{k-q}})-n_{\mathbf{\mathbf{q}\lambda}}+\frac{\delta f_{\mathbf{\mathbf{q}-\mathbf{k}}}}{\delta f_{\mathbf{k}}}(n_{B}(E_{\mathbf{k}})-n_{\mathbf{q}\lambda})\right].\label{eq:tau_t^nc}
\end{align}
 For the magnon number conserving process the derivation is similar
and we find
\begin{align}
\frac{1}{\tau_{\mathbf{k},t}^{c}} & =\frac{\pi\hbar}{mN}\sum_{\mathbf{q}\lambda}\frac{\vert\Gamma_{\mathbf{k},\mathbf{\mathbf{k}-\mathbf{q}},\lambda}^{\bar{b}b}\vert^{2}}{\varepsilon_{\mathbf{q}\lambda}}\Bigg[\delta(E_{\mathbf{k}}-E_{\mathbf{\mathbf{k}-\mathbf{q}}}+\varepsilon_{\mathbf{q}\lambda})\nonumber \\
\times & \left(n_{\mathbf{q}\lambda}-n_{B}(E_{\mathbf{k-q}})-\frac{\delta f_{\mathbf{k}-\mathbf{q}}}{\delta f_{\mathbf{k}}}(n_{B}(E_{\mathbf{k}})+n_{\mathbf{\mathbf{q}\lambda}}+1)\right)\nonumber \\
+ & \delta(E_{\mathbf{k}}-E_{\mathbf{\mathbf{k}-\mathbf{q}}}-\varepsilon_{\mathbf{q}\lambda})\nonumber \\
\times & \left(1+n_{B}(E_{\mathbf{k-q}})+n_{\mathbf{\mathbf{q}\lambda}}+\frac{\delta f_{\mathbf{\mathbf{k}-\mathbf{q}}}}{\delta f_{\mathbf{k}}}(n_{B}(E_{\mathbf{k}})-n_{\mathbf{q}\lambda})\right)\Bigg],\nonumber \\
\label{eq:tau_t^c}
\end{align}
with interaction vertex $\Gamma_{\mathbf{k},\mathbf{k'},\lambda}^{\bar{b}b}$
given by Eq.~(\ref{eq:Gamma_c}). Due to the $\delta f_{\mathbf{\mathbf{k}-\mathbf{q}}}/\delta f_{\mathbf{k}}$
term this is an integral equation. It can be solved iteratively to
generate a geometric series referred to as vertex correction in diagrammatic
theories. By simply disregarding the in-scattering with terms $\delta f_{\mathbf{\mathbf{k}-\mathbf{q}}}/\delta f_{\mathbf{k}}$
the transport lifetime reduces to the the quasi-particle lifetime
of the self-energy. We leave the general solution of this integral
equation for future work, but argue in Sec.~\ref{subsec:results magnon-transport-lifetime}
that the vertex corrections are not important in our regime of interest.

\section{Numerical results\label{sec:Numerical-results}}

\subsection{Magnon decay rate}

In the following we present and analyze our results for the magnon
decay rates in YIG. We first consider the case of vanishing effective
magnetic field ($B=0$) and discuss the magnetic field dependence
in Sec.~\ref{subsec:Magnetic-field-dependence}. Since our model
is only valid in the long-wavelength ($k<8\text{\texttimes}10^{8}\;\mathrm{m}^{-1}$)
and low-temperature ($T\lesssim100\;\mathrm{K}$) regime, we focus
first on $T=50\;\mathrm{K}$ and discuss the temperature dependence
in Sec.~\ref{subsec:Temperature-dependence}. 

\begin{figure}
\begin{centering}
\includegraphics{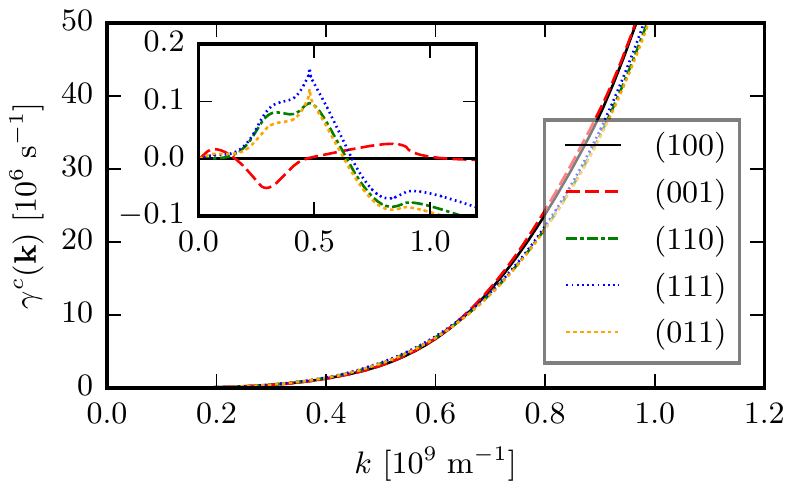}
\par\end{centering}
\caption{Magnon decay rate in YIG due to magnon-phonon interactions for magnons
propagating along various directions at $T=50\;\mathrm{K}$ and $B=0$.\label{fig:gamma c}
We denote the propagation direction by $(lmn)$, i.e. $l\mathbf{e}_{x}+m\mathbf{e}_{y}+n\mathbf{e}_{z}$.
The inset shows the relative deviation $\delta\gamma^{c}/\gamma^{c}$
from the (100) direction.}
\end{figure}

In Fig.~\ref{fig:gamma c} we show the magnon number conserving decay
rate $\gamma^{c}(\mathbf{k})$, which is on the displayed scale dominated
by the exchange-mediated magnon-phonon interaction and is isotropic
for long-wavelength magnons. 

\begin{figure}
\begin{centering}
\includegraphics{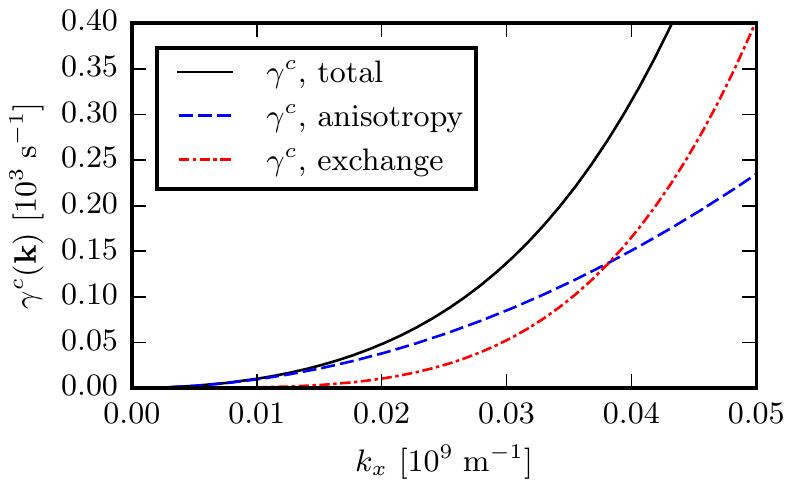}
\par\end{centering}
\caption{Comparison of the contributions from exchange-mediated and anisotropy-mediated
magnon-phonon interactions to the magnon number conserving scattering
rate $\gamma^{c}$ at $T=50\;\mathrm{K}$ and $B=0$.\label{fig:Comparison}}
\end{figure}

In Fig.~\ref{fig:Comparison} we compare the contribution from the
exchange-mediated magnon-phonon interaction ($\gamma^{c}\sim k^{4}$)
and from the anisotropy-mediated magnon-phonon interaction ($\gamma^{c}\sim k^{2}$).
We observe a cross-over at $k\approx4\times10^{7}\;\mathrm{m}^{-1}$:
for much smaller wave numbers, the exchange contribution can be disregarded
and for larger wave numbers the exchange contribution becomes dominant.

\begin{figure}
\begin{centering}
\includegraphics{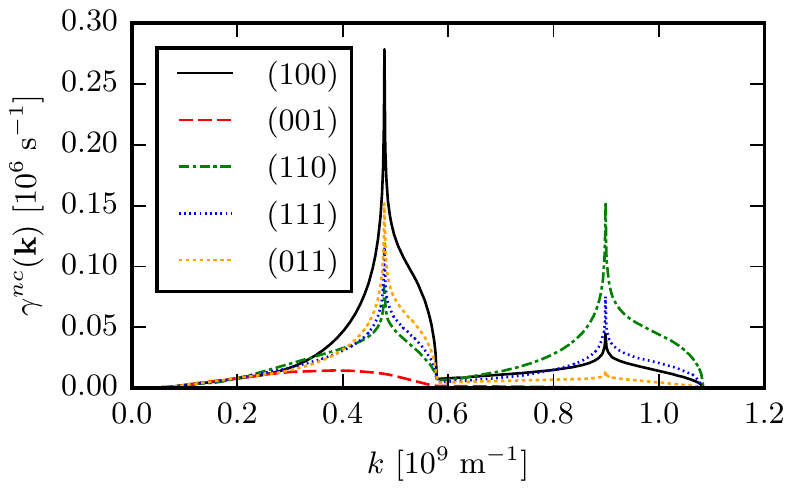}
\par\end{centering}
\caption{Magnon decay rate in YIG due to magnon number non-conserving magnon-phonon
interactions for magnons propagating along various directions at $T=50\;\mathrm{K}$
and $B=0$.\label{fig:gamma nc}}
\end{figure}

The magnon number non-conserving decay rate $\gamma^{nc}$ in Fig.~\ref{fig:gamma nc}
is much smaller than the magnon-conserving one. This is consistent
with the low magnetization damping of YIG, i.e. the magnetization
is long-lived. We observe divergent peaks at the crossing points (shown
in Fig.~\ref{fig:phonon dispersion}) with the exception of the (001)
direction. These divergences occur when magnons and phonons are degenerate
at $k=0.48\times10^{9}\;\mathrm{m}^{-1}$ ($1.2\;\mathrm{meV}$) and
$k=0.9\times10^{9}\;\mathrm{m}^{-1}$ ($4.3\;\mathrm{meV}$), respectively,
at which the Boltzmann formalism does not hold; a treatment in the
magnon-polaron basis \cite{Flebus2017} or a broadening parameter
\cite{Schmidt2018} would get rid of the singular behavior. The divergences
are also suppressed by arbitrarily small effective magnetic fields
(see Sec.~\ref{subsec:Magnetic-field-dependence}). There are no
peaks along the (001) direction because in the (001) direction the
vertex function $V_{\mathbf{q},\lambda}$ (see Eq.~(\ref{eq:V}))
vanishes for $\mathbf{q}=(0,0,k_{z})$. For $k>\hbar c_{l}/(D(\sqrt{8}-2))=1.085\times10^{9}\;\mathrm{m}^{-1}$
the decay rate $\gamma^{nc}$ vanishes because the decay process does
not conserve energy ($\delta(E_{\mathbf{k}}+E_{\mathbf{q}-\mathbf{k}}-\varepsilon_{\mathbf{q}\lambda})=0$).

\subsection{Temperature dependence\label{subsec:Temperature-dependence}}

\begin{figure}
\begin{centering}
\includegraphics{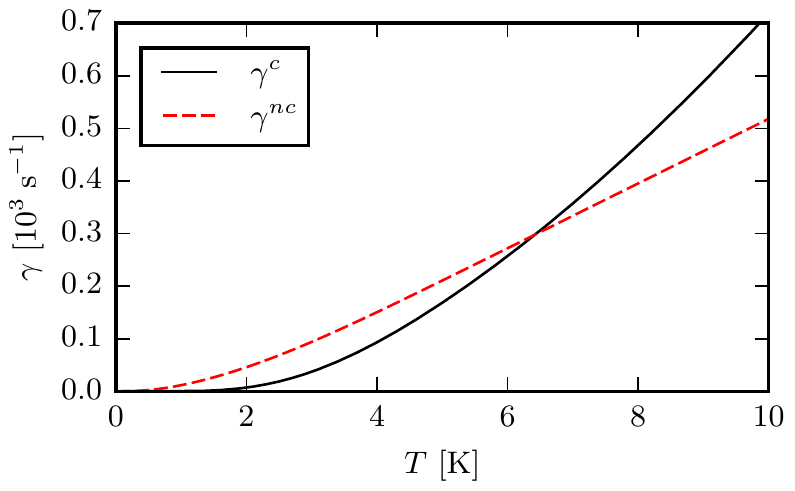}
\par\end{centering}
\caption{Temperature dependence of the magnon decay rates $\gamma^{nc}$ and
$\gamma^{c}$ at $B=0$, $k_{x}=10^{8}\;\mathrm{m}^{-1}$ and $k_{y}=k_{z}=0$,
i.e. along (100).\label{fig: T-dependence 1}}
\end{figure}

Above we focused on $T=50\;\mathrm{K}$ and explained that we expect
a linear temperature dependence of the magnon decay rates at high,
but not low temperatures. Fig.~\ref{fig: T-dependence 1} shows our
results for the temperature dependence at $k_{x}=10^{8}\;\mathrm{m}^{-1}$.
Deviations from the linear dependence at low temperatures occurs when
quantum effects set in, i.e. the Rayleigh-Jeans distribution does
not hold anymore, 
\begin{equation}
\frac{1}{e^{\varepsilon/(k_{B}T)}-1}\not\approx\frac{k_{B}T}{\varepsilon}.
\end{equation}

\subsection{Magnetic field dependence\label{subsec:Magnetic-field-dependence}}

\begin{figure}
\begin{centering}
\includegraphics{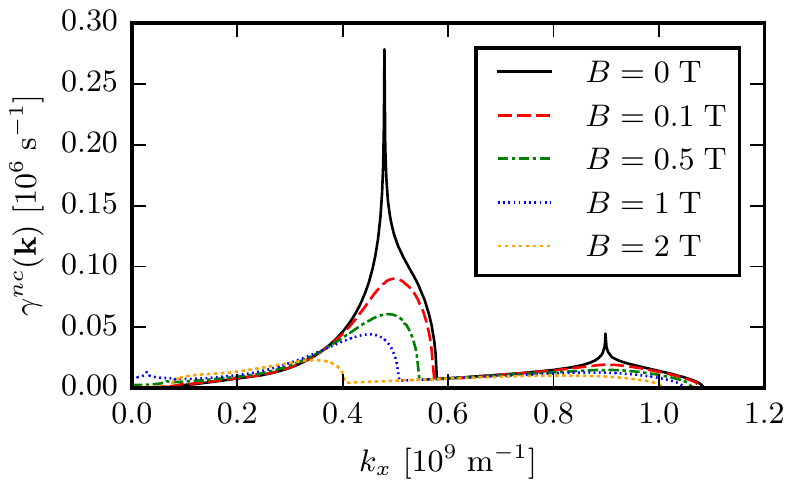}
\par\end{centering}
\caption{Magnetic field dependence of the magnon number non-conserving magnon
decay rate in YIG at $T=50\;\mathrm{K}$ with magnon momentum along
(100).\label{fig:gamma_nc B}}
\end{figure}

The numerical results presented above are for a mono-domain magnet
in the limit of small applied magnetic fields. A finite magnetic field
$B$ along the magnetization direction induces an energy gap $g\mu_{B}B$
in the magnon dispersion, which shifts the positions of the magnon-phonon
crossing points to longer wavelengths. The magnetic field suppresses
the (unphysical) sharp peaks at the crossing points (see Fig.~\ref{fig:gamma_nc B})
that are caused by the divergence of the Planck distribution function
for a vanishing spin wave gap.

\begin{figure}
\begin{centering}
\includegraphics{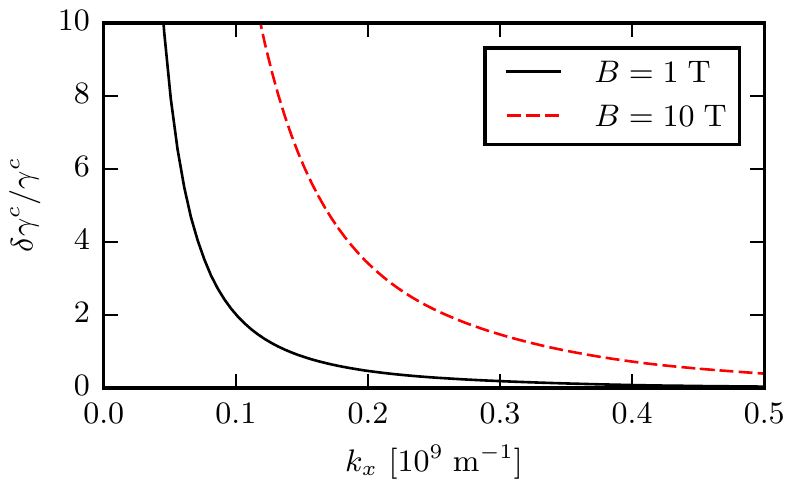}
\par\end{centering}
\caption{Relative deviation $\delta\gamma^{c}/\gamma^{c}$ from the $B=0$
result of the magnon number conserving magnon decay rate in YIG at
$T=50\;\mathrm{K}$ with magnon momentum along (100).\label{fig:gamma_c B}}
\end{figure}

In the magnon number conserving magnon-phonon interactions, the magnetic
field dependence cancels in the delta function and enters only in
the Bose function via $n_{B}$ (magnetic freeze-out). Fig.~\ref{fig:gamma_c B}
shows that the magnetic field mainly affects magnons with energies
$\lesssim2g\mu_{B}B=0.23(B/\mathrm{T})\;\mathrm{meV}$.

\begin{figure}
\begin{centering}
\includegraphics{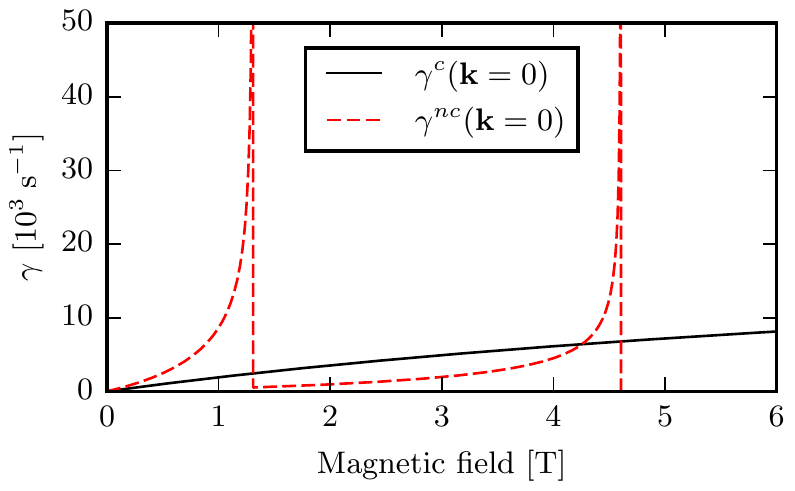}
\par\end{centering}
\caption{Magnetic field dependence of the magnon decay rates in YIG at $\mathbf{k}=0$
and $T=50\;\mathrm{K}$.\label{fig:gamma_field}}

\end{figure}

As shown in Fig. \ref{fig:gamma_field} the magnon decay by phonons
does not vanish for the $\mathbf{k}=0$ Kittel mode, but only in the
presence of a spin wave gap $E_{0}=g\mu_{B}B$. Both magnon conserving
and non-conserving scattering processes contribute. The divergent
peaks at $B\approx1.3\;\mathrm{T}$ and $B\approx4.6\;\mathrm{T}$
in $\gamma^{nc}$ are caused by energy and momentum conservation in
the two-magnon-one-phonon scattering process,

\begin{equation}
\delta(E_{\mathbf{k}=0}+E_{\mathbf{q}}-\varepsilon_{\mathbf{q}\lambda})=\delta(2g\mu_{B}B+E_{ex}q^{2}a^{2}-\hbar c_{\lambda}q),
\end{equation}
when the gradient of the argument of the delta function vanishes,
\begin{equation}
\boldsymbol{\nabla}_{\mathbf{q}}(E_{\mathbf{k}=0}+E_{\mathbf{q}}-\varepsilon_{\mathbf{q}\lambda})=0,
\end{equation}
i.e., the two-magnon energy $E_{\mathbf{k}=0}+E_{\mathbf{q}}$ touches
either the transverse or longitudinal phonon dispersion $\varepsilon_{\mathbf{q}\lambda}$.
The total energy of the two magnons is equivalent to the energy of
a single magnon with momentum $q$ but in a field $2B$, resulting
in the divergence at fields that are half of those for the magnon-polaron
observed in the spin Seebeck effect \cite{Flebus2017,Schmidt2018}.
The two-magnon touching condition can be satisfied in all directions
of the phonon momentum $\mathbf{q}$, which therefore contributes
to the magnon decay rate when integrating over the phonon momentum
$\mathbf{q}$. For $\mathbf{k}\neq0$ this two-magnon touching condition
can only be fulfilled for phonons along a particular direction and
the divergence is suppressed.

The magnon decay rate is related to the Gilbert damping $\alpha_{\mathbf{k}}$
as $\hbar\gamma_{\mathbf{k}}=2\alpha_{\mathbf{k}}E_{\mathbf{k}}$
\cite{Bender2014}. We find that phonons contribute only weakly to
the Gilbert damping, $\alpha_{0}^{nc}=\hbar\gamma_{0}^{nc}/(2E_{0})\sim10^{-8}$
at $T=50\;\mathrm{K}$, which is much smaller than the total Gilbert
damping $\alpha\sim10^{-5}$ in YIG, but the peaks at $1.3\;\mathrm{T}$
and $4.6\;\mathrm{T}$ might be observable. The phonon contribution
to the Gilbert damping scales linearly with temperature, so is twice
as large at 100 K. At low temperatures ($T\lesssim100\;\mathrm{K}$)
Gilbert damping in YIG has been found to be caused by two-level systems
\cite{Tabuchi2014} and impurity scattering \cite{Maier-Flaig2017},
while for higher temperatures magnon-phonon \cite{Kasuya1961} and
magnon-magnon scattering involving optical magnons \cite{Cherepanov1993}
have been proposed to explain the observed damping. Enhanced damping
as a function of magnetic field at higher temperatures might reveal
other van Hove singularities in the joint magnon-phonon density of
states.

\subsection{Magnon transport lifetime\label{subsec:results magnon-transport-lifetime}}

We do not attempt a full solution of the integral equations (\ref{eq:tau_t^nc})
and (\ref{eq:tau_t^c}) for the transport lifetime. However, we can
still estimate its effect by the observation that the ansatz $\tau_{\mathbf{k},t}^{-1}\sim k^{n}$
can be an approximate solution of the Boltzmann equation with in-scattering.

\begin{figure}
\begin{centering}
\includegraphics{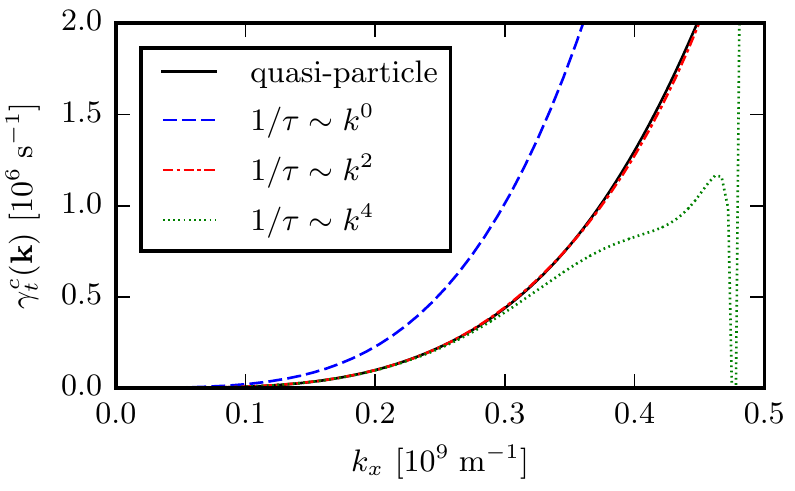}
\par\end{centering}
\caption{Inverse of the magnon transport lifetime in YIG (with magnon momentum
along (100)) due to magnon number conserving magnon-phonon interactions
at $T=50\;\mathrm{K}$ and $B=0$ for magnons along the (100) direction.\label{fig:transport c}}
\end{figure}

Our results for the magnon number conserving interaction are shown
in Fig.~\ref{fig:transport c} (for $\boldsymbol{\nabla}T=0$ and
finite $\boldsymbol{\nabla}\mu||\mathbf{e}_{x}$), where $\gamma_{t}=\tau_{t}^{-1}$.
We consider the cases $n=0,2,4$, where $n=0$ or $\tau_{\mathbf{k},t}=\mathrm{const}.$
would be the solution for a short-range scattering potential. For
very long wavelengths ($k\lesssim4\times10^{7}\;\mathrm{m}^{-1}$)
the inverse quasi-particle lifetime $\tau_{\mathbf{k},qp}^{-1}\sim k^{2}$
and for shorter wavelengths $\tau_{\mathbf{k},qp}^{-1}\sim k^{4}$.
$n=2$ is a self-consistent solution only for very small $k\lesssim4\times10^{7}\;\mathrm{m}^{-1}$,
while $\tau_{\mathbf{k},qp}^{-1}\sim k^{4}$ is a good ansatz up to
$k\lesssim0.3\times10^{9}\;\mathrm{m}^{-1}$. We see that the transport
lifetime approximately equals the quasi-particle lifetime in the regime
of the validity of the $n=4$ power law. 

\begin{figure}
\begin{centering}
\includegraphics{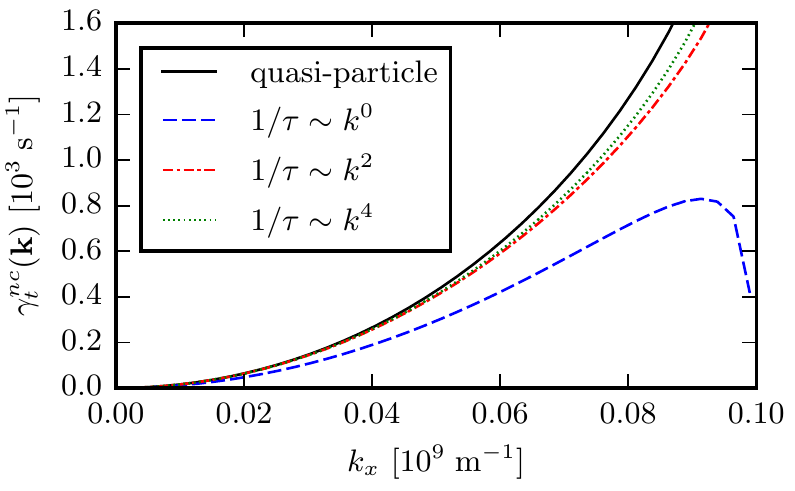}
\par\end{centering}
\caption{Inverse of the magnon transport lifetime in YIG (with magnon momentum
along (100)) due to magnon number non-conserving interactions at $T=50\;\mathrm{K}$
and $B=0$.\label{fig:transport nc}}
\end{figure}

For the magnon number non-conserving processes in Fig.~\ref{fig:transport nc}
the quasi-particle lifetime behaves as $\tau_{\mathbf{k},qp}^{-1}\sim k^{2}$.
The ansatz $n=2$ turns out to be self-consistent and we see deviations
of the transport lifetime from the quasi-particle lifetime for $k\gtrsim5\times10^{7}\;\mathrm{m}^{-1}$.
The plot only shows our results for $k<1\times10^{8}\;\mathrm{m}^{-1}$
because our assumption of an isotropic lifetime is not valid for higher
momenta in this case.

We conclude that for YIG in the long-wavelength regime the magnon
transport lifetime (due to magnon-phonon interactions) should be approximately
the same as the quasi-particle lifetime, but deviations at shorter
wavelengths require more attention.


\section{Summary and conclusion\label{sec:Summary}}

We calculated the decay rate of magnons in YIG induced by magnon-phonon
interactions in the long-wavelength regime ($k\lesssim1\times10^{9}\;\mathrm{m}^{-1}$).
Our model takes only the acoustic magnon and phonon branches into
account and is therefore valid at low to intermediate temperatures
($T\lesssim100\;\mathrm{K}$). The exchange-mediated magnon-phonon
interaction has been recently identified as a crucial contribution
to the overall magnon-phonon interaction in YIG at high temperatures
\cite{Cornelissen2016,Maehrlein2017,Liu2017}. We emphasize that its
coupling strength can be derived from experimental values of the magnetic
Grüneisen parameter $\Gamma_{m}=\partial\ln T_{C}/\partial\ln V$
\cite{Bloch1966,Kamilov1998}. In previous works this interaction
has been either disregarded \cite{Rueckriegel2014}, underestimated
\cite{Maehrlein2017,Shklovskij2018}, or overestimated \cite{Cornelissen2016}.

In the ultra-long-wavelength regime the wave vector dependent magnon
decay rate $\gamma(\mathbf{k})$ is determined by the anisotropy-mediated
magnon-phonon interaction with $\gamma(\mathbf{k})\sim k^{2}$, while
for shorter wavelengths $k\gtrsim4\times10^{7}\;\mathrm{m}^{-1}$
the exchange-mediated magnon-phonon interaction becomes dominant,
which scales as $\gamma(\mathbf{k})\sim k^{4}$. The magnon number
non-conserving processes are caused by spin-orbit interaction, i.e.,
the anisotropy-mediated magnon-phonon interaction, and are correspondingly
weak.

In a finite magnetic field the average phonon scattering contribution,
from the mechanism under study, to the Gilbert damping of the $k=0$
macrospin Kittel mode is about three orders of magnitude smaller than
the best values for the Gilbert damping $\alpha\sim10^{-5}$. However,
we predict peaks at $1.3\;\mathrm{T}$ and $4.6\;\mathrm{T}$, that
may be experimentally observable in high-quality samples.

The magnon transport lifetime, which is given by the balance between
in- and out-scattering in the Boltzmann equation, is in the long-wavelength
regime approximately the same as the quasi-particle lifetime. However,
the magnon quasi-particle and transport lifetime differ more significantly
at shorter wavelengths. A theory for magnon transport at room temperature
should therefore include the ``vertex corrections''.

A full theory of magnon transport at high temperature requires a method
that takes the full dispersion relations of acoustic and optical phonons
and magnons into account. This would also require a full microscopic
description of the magnon-phonon interaction, since the magnetoelastic
energy used here only holds in the continuum limit. 
\begin{acknowledgments}
N. V-S thanks F. Mendez for useful discussions. This work is part
of the research program of the Stichting voor Fundamenteel Onderzoek
der Materie (FOM), which is financially supported by the Nederlandse
Organisatie voor Wetenschappelijk Onderzoek (NWO) as well as a Grant-in-Aid
for Scientific Research on Innovative Area, \textquotedblright Nano
Spin Conversion Science\textquotedblright{} (Grant No. 26103006),
CONICYT-PCHA/Doctorado Nacional/2014-21140141, Fondecyt Postdoctorado
No. 3190264, and Fundamental Research Funds for the Central Universities.
\end{acknowledgments}

\appendix

\section{Long-wavelength approximation\label{sec:Longwavelength-approximation}}

\begin{figure*}
\begin{centering}
\includegraphics{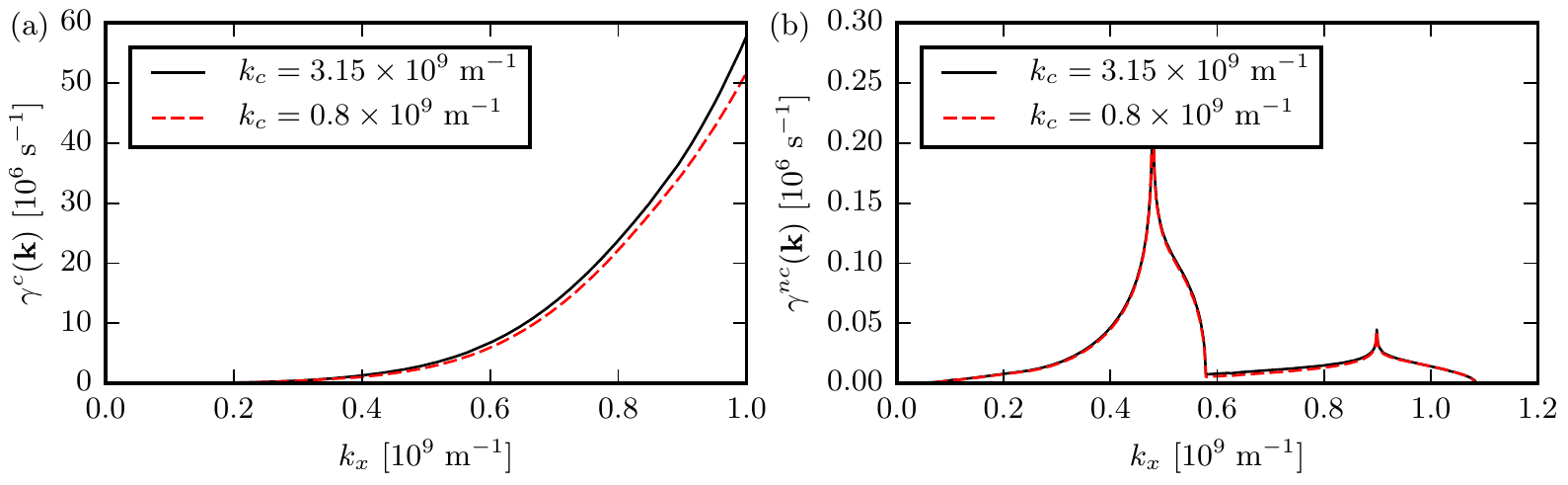}
\par\end{centering}
\caption{Dependence the magnon decay rate along (100) on the high magnon momentum
cut-off $k_{c}$ for the (a) magnon number conserving ($\gamma^{c}$)
and (b) non-conserving ($\gamma^{nc}$) contributions at $T=50\;\mathrm{K}$
and $B=0$.\label{fig:cutoff k}}
\end{figure*}

The theory is designed for magnons with momentum $k<0.8\times10^{9}\;\mathrm{m}^{-1}$
and phonons with momentum $q<2.5\times10^{9}\;\mathrm{m}^{-1}$ (corresponding
to phonon energies/frequencies $\leq$$12\;\mathrm{meV}$/$3\;\mathrm{THz}$),
but relies on high-momentum cut-off parameters $k_{c}$ because of
the assumption of quadratic/linear dispersion of magnon/phonons. We
see in Fig.~\ref{fig:cutoff k} that the scattering rates only weakly
depend on $k_{c}$. 

\begin{figure*}
\begin{centering}
\includegraphics{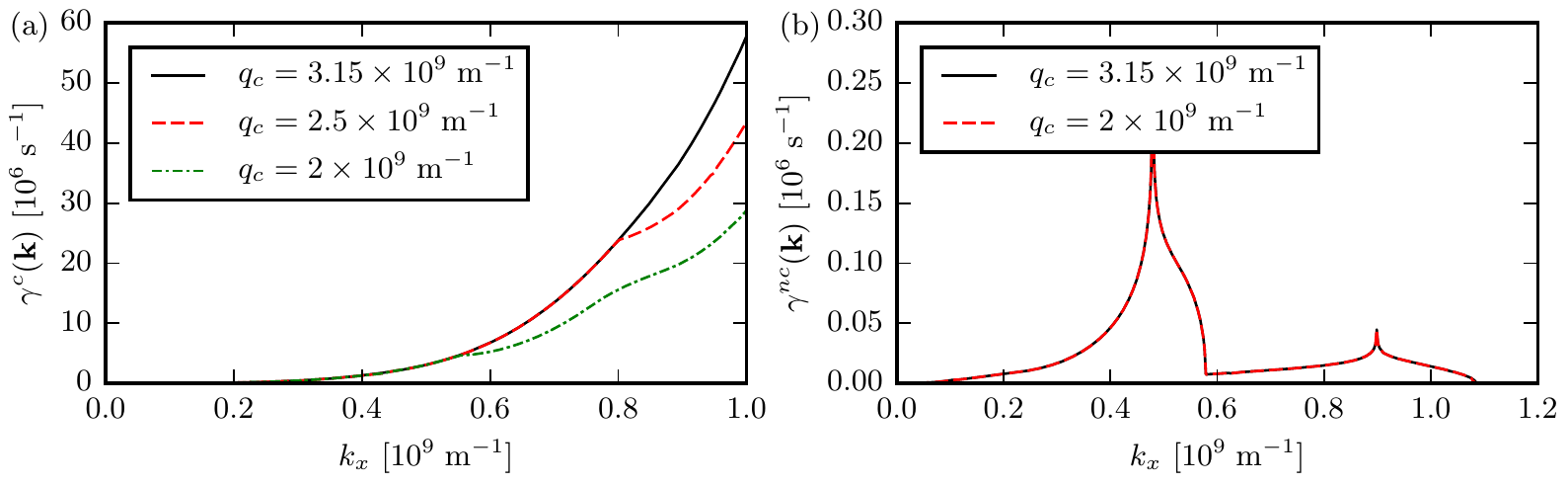}
\par\end{centering}
\caption{Dependence the magnon decay rate along (100) on the high phonon momentum
cut-off $q_{c}$ for the (a) magnon number conserving ($\gamma^{c}$)
and (b) non-conserving ($\gamma^{nc}$) contributions at $T=50\;\mathrm{K}$
and $B=0$.\label{fig:cutoff q}}
\end{figure*}

The dependence of the scattering rate on the phonon momentum cut-off
$q_{c}$ is shown in Fig.~\ref{fig:cutoff q}. $q_{c}=3.15\times10^{9}\;\mathrm{m}^{-1}$
corresponds to an integration over the whole Brillouin zone, approximated
by a sphere. From these considerations we estimate that the long-wavelength
approximation is reliable for $k\lesssim8\times10^{8}\;\mathrm{m}^{-1}$.
Optical phonons (magnons) that are thermally excited for $T\apprge100\;\mathrm{K}\,(300\;\mathrm{K})$
are not considered here.

\section{Second order magnetoelastic coupling\label{sec:Second-order-magnetoelastic}}

The magnetoelastic energy is usually expanded only to first order
in the displacement fields. Second order terms can become important
e.g. when the first order terms vanish. This is the case for one-magnon
two-phonon scattering processes. The first order term
\begin{equation}
\sum_{\mathbf{q}\lambda}\left[\Gamma_{\mathbf{q}\lambda}b_{-\mathbf{q}}X_{\mathbf{q}\lambda}+\Gamma_{-\mathbf{q}\lambda}^{*}b_{\mathbf{q}}^{\dagger}X_{\mathbf{q}\lambda}\right]
\end{equation}
only contributes when phonon and magnon momenta and energies cross,
giving rise to magnon polaron modes \cite{Flebus2017}. In other areas
of reciprocal space the second order term should therefore be considered.
Eastman \cite{Eastman1966,Eastman1966b} derived the second-order
magnetoelastic energy and determined the corresponding coupling constants
for YIG. In momentum space, the relevant contribution to the Hamiltonian
is of the form
\begin{align}
\mathcal{H}_{2p1m} & =\frac{1}{\sqrt{N}}\sum_{\mathbf{k},\mathbf{q}_{1},\lambda_{1},\mathbf{q}_{2},\lambda_{2}}\nonumber \\
 & \left(\delta_{\mathbf{q}_{1}+\mathbf{q}_{2}+\mathbf{k},0}\Gamma_{\mathbf{q}_{1}\lambda_{1},\mathbf{q}_{2}\lambda_{2}}^{b}X_{\mathbf{q}_{1}\lambda_{1}}X_{\mathbf{q}_{2}\lambda_{2}}b_{\mathbf{k}}\right.\nonumber \\
 & \left.+\delta_{\mathbf{q}_{1}+\mathbf{q}_{2}-\mathbf{k},0}\Gamma_{\mathbf{q}_{1}\lambda_{1},\mathbf{q}_{2}\lambda_{2}}^{\bar{b}}X_{\mathbf{q}_{1}\lambda_{1}}X_{\mathbf{q}_{2}\lambda_{2}}b_{\mathbf{k}}^{\dagger}\right),
\end{align}
where the interaction vertices are symmetrized,
\begin{equation}
\Gamma_{\mathbf{q}_{1}\lambda_{1},\mathbf{q}_{2}\lambda_{2}}^{b}=\frac{1}{2}\left(\tilde{\Gamma}_{\mathbf{q}_{1}\lambda_{1},\mathbf{q}_{2}\lambda_{2}}^{b}+\tilde{\Gamma}_{\mathbf{q}_{2}\lambda_{2},\mathbf{q}_{1}\lambda_{1}}^{b}\right),
\end{equation}
and obey
\begin{equation}
\Gamma_{\mathbf{q}_{1}\lambda_{1},\mathbf{q}_{2}\lambda_{2}}^{b}=\left(\Gamma_{-\mathbf{q}_{1}\lambda_{1},-\mathbf{q}_{2}\lambda_{2}}^{\bar{b}}\right)^{*}.
\end{equation}
The non-symmetrized vertex function is
\begin{align}
\tilde{\Gamma}_{\mathbf{q}_{1}\lambda_{1},\mathbf{q}_{2}\lambda_{2}}^{b} & =\frac{1}{a^{2}\sqrt{2S}}\left[B_{144}\left(iI_{1}-I_{1,x\leftrightarrow y}\right)\right.\nonumber \\
 & +B_{155}\left(iI_{2}-I_{2,x\leftrightarrow y}\right)\nonumber \\
 & +\left.B_{456}\left(iI_{3}-I_{3,x\leftrightarrow y}\right)\right],\label{eq:two-phonon interaction}
\end{align}
with
\begin{align}
I_{1} & =a^{2}e_{\mathbf{q}_{1}\lambda_{1}}^{x}q_{1}^{x}\left[e_{\mathbf{q}_{2}\lambda_{2}}^{y}q_{2}^{z}+e_{\mathbf{q}_{2}\lambda_{2}}^{z}q_{2}^{y}\right],\\
I_{2} & =a^{2}\left[e_{\mathbf{q}_{1}\lambda_{1}}^{y}q_{1}^{y}+e_{\mathbf{q}_{1}\lambda_{1}}^{z}q_{1}^{z}\right]\nonumber \\
 & \times\left[e_{\mathbf{q}_{2}\lambda_{2}}^{y}q_{2}^{z}+e_{\mathbf{q}_{2}\lambda_{2}}^{z}q_{2}^{y}\right],\\
I_{3} & =a^{2}\left[e_{\mathbf{q}_{1}\lambda_{1}}^{x}q_{1}^{z}+e_{\mathbf{q}_{1}\lambda_{1}}^{z}q_{1}^{x}\right]\nonumber \\
 & \times\left[e_{\mathbf{q}_{2}\lambda_{2}}^{x}q_{2}^{y}+e_{\mathbf{q}_{2}\lambda_{2}}^{y}q_{2}^{x}\right],
\end{align}
and $x\leftrightarrow y$ denotes an exchange of $x$ and $y$. The
relevant coupling constants in YIG are \cite{Eastman1966b,Eastman1966}
\begin{align}
B_{144} & =-6\pm48\;\mathrm{meV},\\
B_{155} & =-44\pm6\;\mathrm{meV},\\
B_{456} & =-32\pm8\;\mathrm{meV}.
\end{align}

\begin{figure}
\begin{centering}
\includegraphics[scale=0.4]{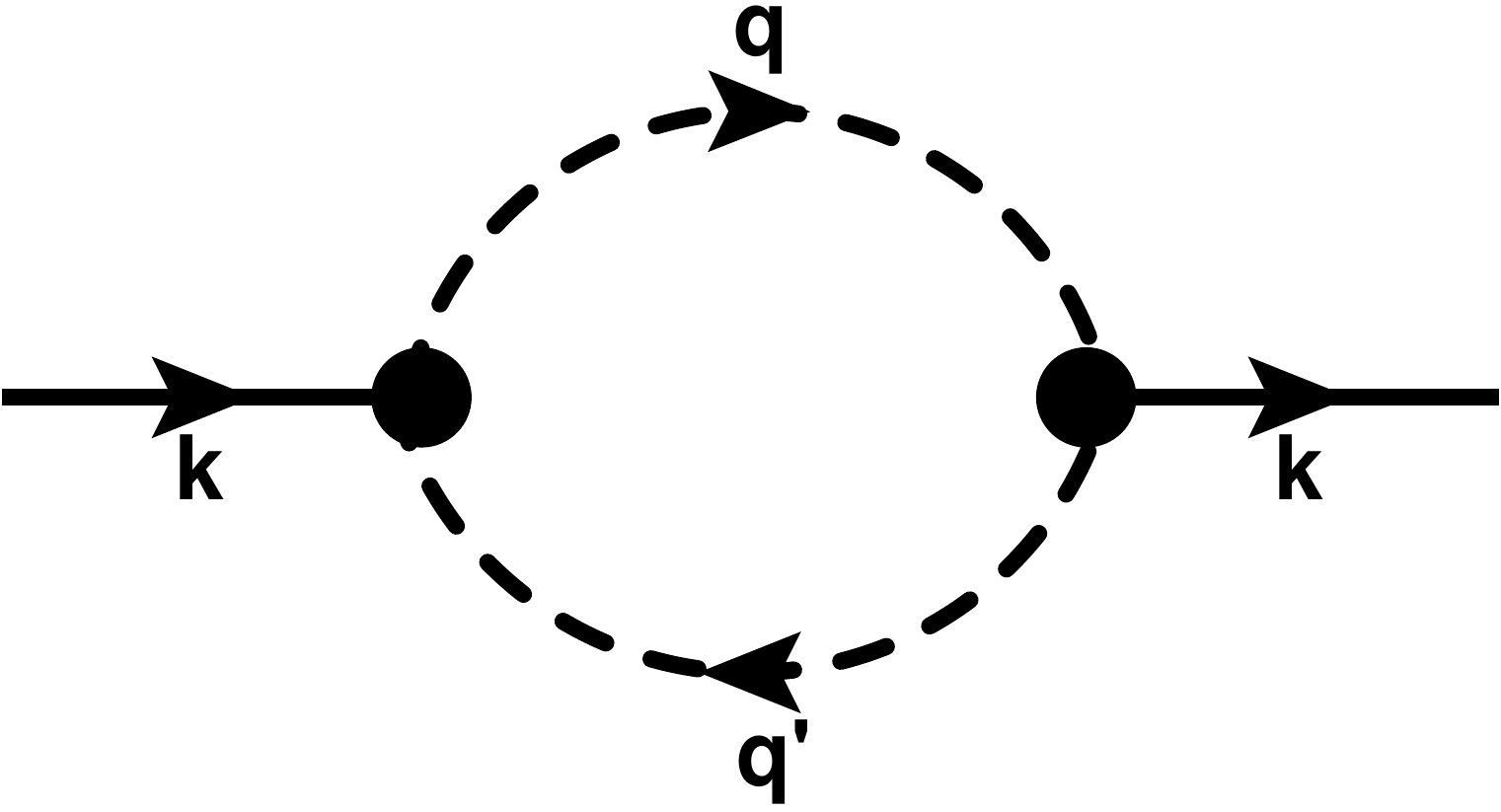}
\par\end{centering}
\caption{Feynman diagram representing the self-energy Eq.~(\ref{eq:self-energy-2})
due to one-magnon-two-phonon processes.\label{fig:self-energy-2}}
\end{figure}

The magnon self-energy (see Fig.~\ref{fig:self-energy-2}) reads
\begin{align}
\Sigma_{2p1m}(\mathbf{k},i\omega) & =-\frac{2}{N}\sum_{\mathbf{q}_{1},\lambda_{1},\mathbf{q}_{2},\lambda_{2}}\frac{1}{\beta}\sum_{\Omega}\delta_{\mathbf{q}_{1}+\mathbf{q}_{2}+\mathbf{k},0}\nonumber \\
 & \times\left|\Gamma_{\mathbf{q}_{1}\lambda_{1},\mathbf{q}_{2}\lambda_{2}}^{b}\right|^{2}F_{\lambda_{1}}(\mathbf{q}_{1},\Omega)F_{\lambda_{2}}(\mathbf{q}_{2},-\Omega-\omega).\label{eq:self-energy-2}
\end{align}
with phonon propagator
\begin{equation}
F_{\lambda}(\mathbf{q},\Omega)=\frac{\hbar^{2}}{m}\frac{1}{\hbar^{2}\Omega^{2}+\varepsilon_{\mathbf{q}\lambda}^{2}}.
\end{equation}
 and leads to a magnon decay rate 
\begin{align}
\gamma_{2p}^{nc}(\mathbf{k}) & =-\frac{2}{\hbar}\mathrm{Im}\Sigma_{2p1m}(\mathbf{k},i\omega\to E_{\mathbf{k}}/\hbar+i0^{+})\nonumber \\
 & =\frac{\pi\hbar^{3}}{m^{2}N}\sum_{\mathbf{q}_{1},\lambda_{1},\mathbf{q}_{2},\lambda_{2}}\delta_{\mathbf{q}_{1}+\mathbf{q}_{2}+\mathbf{k},0}\frac{1}{\varepsilon_{1}\varepsilon_{2}}\left|\Gamma_{\mathbf{q}_{1}\lambda_{1},\mathbf{q}_{2}\lambda_{2}}^{b}\right|^{2}\nonumber \\
 & \times\left\{ 2\delta\left(E_{\mathbf{k}}+\varepsilon_{1}-\varepsilon_{2}\right)\left[n_{1}-n_{2}\right]\right.\nonumber \\
 & \left.+\delta\left(E_{\mathbf{k}}-\varepsilon_{1}-\varepsilon_{2}\right)\left[1+n_{1}+n_{2}\right]\right\} ,\label{eq:mdr}
\end{align}
where 
\begin{align}
n_{1} & =n_{B}\left(\varepsilon_{\mathbf{q}_{1}\lambda_{1}}\right),\;n_{2}=n_{B}\left(\varepsilon_{\mathbf{q}_{2}\lambda_{2}}\right),\\
\varepsilon_{1} & =\varepsilon_{\mathbf{q}_{1}\lambda_{1}},\;\varepsilon_{2}=\varepsilon_{\mathbf{q}_{2}\lambda_{2}}.
\end{align}
The first term in curly brackets on the right-hand-side of Eq. (\ref{eq:mdr})
describes annihilation and creation of a phonon as a sum of out-scattering
minus in-scattering contributions,
\begin{equation}
n_{1}(1+n_{2})-(1+n_{1})n_{2}=n_{1}-n_{2},
\end{equation}
while the second term can be understood in terms of out-scattering
by the creation of two phonons and the in-scattering by annihilation
of two phonons,

\begin{equation}
(1+n_{1})(1+n_{2})-n_{1}n_{2}=1+n_{1}+n_{2}.
\end{equation}
For this one-magnon-two-phonon process the quasi-particle and the
transport lifetimes are the same,
\begin{equation}
\tau_{t}=\tau_{qp},
\end{equation}
since this process involves only a single magnon that is either annihilated
or created. The collision integral is then independent of the magnon
distribution of other magnons and the transport lifetime reduces to
the quasi-particle lifetime.

\begin{figure}
\begin{centering}
\includegraphics{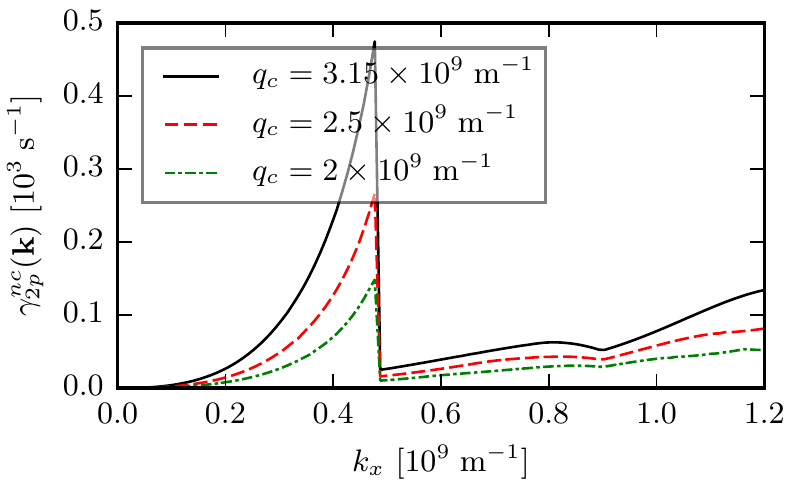}
\par\end{centering}
\caption{Two-phonon contribution to the magnon number non-conserving magnon
scattering rate with magnon momentum along (100) for different values
of the phonon momentum cutoff $q_{c}$ at $T=50\;\mathrm{K}$ and
$B=0$.\label{fig:2p}}
\end{figure}

The two-phonon contribution to the magnon scattering rate in YIG at
$T=50\;\mathrm{K}$ and along (100) direction as shown in Fig.~\ref{fig:2p}
is more than two orders of magnitude smaller than that from one-phonon
processes and therefore disregarded in the main text. The numerical
results depend strongly on the phonon momentum cutoff $q_{c}$, even
in the long-wavelength regime, which implies that the magnons in this
process dominantly interact with short-wavelength, thermally excited
phonons. Indeed, the second order magnetoelastic interaction (\ref{eq:two-phonon interaction})
is quadratic in the phonon momenta, which favors scattering with short-wavelength
phonons. Our long-wavelength approximation therefore becomes questionable
and the results may be not accurate at $T=50\;\mathrm{K}$, but this
should not change the main conclusion that we can disregard these
diagrams.

Our finding that the two-phonon contributions are so small can be
understood in terms of the dimensionful prefactors of the decay rates
(Eqs.~(\ref{eq:gamma_nc_out}-\ref{eq:gamma_nc_in}) and (\ref{eq:mdr})):
The one-phonon decay rate is proportional to $\hbar/(ma^{2})\approx7\times10^{6}\;\mathrm{s^{-1}}$,
while the two-phonon decay rate is proportional to $\hbar^{3}/(m^{2}a^{4}\varepsilon)\approx33\;\mathrm{s}^{-1}$,
where $\varepsilon\approx1\;\mathrm{meV}$ is a typical phonon energy.
The coupling constants for the magnon number non-conserving processes
are $B_{\parallel,\perp}\sim5\;\mathrm{meV}$ while the strongest
two phonon coupling which enhances the two-phonon process by about
a factor 100, but does not nearly compensate the prefactor. The two
phonon process is therefore three orders of magnitudes smaller than
the contribution of the one phonon process. The physical reason appears
to be the large mass density of YIG, i.e. the heavy yttrium atoms.


\section{Numerical integration\label{sec:Numerical-integration}}

The magnon decay rate is given be the weighted density of states

\begin{equation}
I=\int_{BZ}d^{3}q\,f(\mathbf{q})\delta\left(\varepsilon(\mathbf{q})\right),
\end{equation}
that contain the Dirac delta function $\delta(\varepsilon)$ that
can be eliminated to yield

\begin{equation}
I=\sum_{\mathbf{q}_{i}}\int_{A_{i}}d^{2}q\,\frac{f(\mathbf{q})}{\left|\mathbf{\nabla}\varepsilon(\mathbf{q})\right|},
\end{equation}
where the $\mathbf{q}_{i}$ are the zeros of $\varepsilon(\mathbf{q})$
and $A_{i}$ the surfaces inside the Brillouin zone with $\varepsilon(\mathbf{q})=\varepsilon(\mathbf{q}_{i})$.
The calculation these integrals is a standard numerical problem in
condensed matter physics.

For a spherical Brillouin zone of radius $q_{c}$ and spherical coordinates
$(q,\theta,\phi)$,
\begin{equation}
I=\int_{0}^{\pi}d\theta\int_{0}^{2\pi}d\phi\int_{0}^{q_{c}}dq\,q^{2}\sin(\theta)f(q,\theta,\phi)\delta\left(\varepsilon(q,\theta,\phi)\right).
\end{equation}
When $\varepsilon(q_{i},\theta,\phi)=0$ 
\begin{equation}
\delta\left(\varepsilon(q,\theta,\phi)\right)=\sum_{q_{i}\left(\theta,\phi\right)}\frac{\delta\left(q-q_{i}\left(\theta,\phi\right)\right)}{\left|\varepsilon'(q_{i}\left(\theta,\phi\right),\theta,\phi)\right|},
\end{equation}
where $\varepsilon'=\partial\varepsilon/\partial q$ and
\begin{align}
I & =\int_{0}^{\pi}d\theta\int_{0}^{2\pi}d\phi\,\sum_{q_{i}(\theta,\phi)<q_{c}}q_{i}^{2}(\theta,\phi)\sin(\theta)\nonumber \\
 & \phantom{\int_{0}^{\pi}d\theta\int_{0}^{2\pi}d\phi\,\sum_{q_{i}(\theta,\phi)<q_{c}}}\times\frac{f(q_{i}\left(\theta,\phi\right),\theta,\phi)}{\left|\varepsilon'(q_{i}\left(\theta,\phi\right),\theta,\phi)\right|},
\end{align}
which is particularly useful when the zeros of $\varepsilon(q,\theta,\phi)$
can be calculated analytically for linear and quadratic dispersion
relations.

We can also evaluate the integral $I$ fully numerically by broadening
the delta function \cite{Illg2016} e.g. replacing it by a Gaussian
\cite{Illg2016},
\begin{equation}
\delta(\varepsilon)\to\frac{1}{\sqrt{\pi}\sigma}\exp\left(-\frac{\varepsilon^{2}}{\sigma^{2}}\right),
\end{equation}
where $\sigma$ is the broadening parameter. An alternative is the
Lorentzian (Cauchy-Lorentz distribution),
\begin{equation}
\delta(\varepsilon)\to\frac{1}{\pi\sigma}\frac{\sigma^{2}}{\varepsilon^{2}+\sigma^{2}},
\end{equation}
which has fat tails that are helpful in finding the zeros of the delta
function for an adaptive integration grid. Here we use the cubature
package by Steven G. Johnson \cite{Johnson2017}, which implements
an adaptive multidimensional integration algorithm over hyperrectangular
regions \cite{Genz1980,Berntsen1991}.

\bibliographystyle{apsrev4-1}

\end{document}